\documentclass[prb,preprint,superscriptaddress,showpacs]{revtex4}
\usepackage{ae,aecompl}
\usepackage{helvet}
\usepackage[T1]{fontenc}
\usepackage[latin9]{inputenc}
\usepackage{textcomp}
\usepackage{amsmath}
\usepackage{graphicx}
\usepackage{amssymb}
\usepackage{esint}

\makeatletter
\@ifundefined{textcolor}{}
{%
 \definecolor{BLACK}{gray}{0}
 \definecolor{WHITE}{gray}{1}
 \definecolor{RED}{rgb}{1,0,0}
 \definecolor{GREEN}{rgb}{0,1,0}
 \definecolor{BLUE}{rgb}{0,0,1}
 \definecolor{CYAN}{cmyk}{1,0,0,0}
 \definecolor{MAGENTA}{cmyk}{0,1,0,0}
 \definecolor{YELLOW}{cmyk}{0,0,1,0}
 }

\def\ketpsi{\vert \psi \rangle}\def\brapsi{\langle \psi \vert}\def\ketpsi0{\vert \psi_0 \rangle}\def\brapsi0{\langle \psi_0 \vert}

\makeatother

\makeatother

\makeatother

\makeatother

\makeatother

\begin{document}

\title{Superconductor-Insulator transition and energy localization}

\author{M.V. Feigel'man}

\affiliation{L.D.Landau Institute for Theoretical Physics, Kosygin str.2, Moscow
119334, Russia}

\author{L.B. Ioffe}

\affiliation{Center for Materials Theory, Department of Physics and Astronomy,
Rutgers University 136 Frelinghuysen Rd, Piscataway NJ 08854 USA}

\author{M. Mézard}

\affiliation{CNRS, Université Paris-Sud, UMR 8626, LPTMS, Orsay Cedex, F-91405
France}

\date{\today}
\begin{abstract}
We develop an analytical theory for generic disorder-driven quantum
phase transitions. We apply this formalism to the superconductor-insulator
transition and we briefly discuss the applications to the order-disorder
transition in quantum magnets. The effective spin-$\frac{1}{2}$ models
for these transitions are solved in the cavity approximation which
becomes exact on a Bethe lattice with large branching number $K\gg1$
and weak dimensionless coupling $g\ll1$. The characteristic features
of the low temperature phase is a large self-formed inhomogeneity
of the order-parameter distribution near the critical point $K\geq K_{c}(g)$
where the critical temperature $T_{c}$ of the ordering transition
vanishes. We find that the local probability distribution $P(B)$
of the order parameter $B$ has a long power-law tail in the region
where $B$ is much larger than its typical value $B_{0}$. Near the
quantum critical point, at $K\to K_{c}(g)$, the typical value of
the order parameter vanishes exponentially, $B_{0}\propto e^{-C/(K-K_{c}(g))}$
while the spatial scale $N_{inh}$ of the order parameter inhomogeneities
diverges as $(K-K_{c}(g))^{-2}$. In the disordered regime, realized
at $K<K_{c}(g)$ we find actually two distinct phases characterized
by different behavior of relaxation rates. The first phase exists
in an intermediate range of $K^{*}(g)<K<K_{c}(g)$. It has two regimes
of energies: at low excitation energies, $\omega<\omega_{d}(K,g)$,
the many-body spectrum of the model is \textit{discrete}, with zero
level widths, while at $\omega>\omega_{d}$ the level acquire a non-zero
width which is self-generated by the many-body interactions. In this
phase the spin model provides by itself an intrinsic thermal bath.
Another phase is obtained at smaller $K<K^{*}(g)$, where all the
eigenstates are discrete, corresponding to full many-body localization.
These results provide an explanation for the activated behavior of
the resistivity in amorphous materials on the insulating side near
the SI transition and a semi-quantitative description of the scanning
tunneling data on its superconductive side. 
\end{abstract}


\maketitle
\tableofcontents{}

\section{Introduction}

Recently the subject of zero temperature quantum phase transitions
and of the corresponding quantum critical points in translationally
invariant systems got a lot of attention, the theoretical description
of this phenomenon is mostly complete.\cite{Sachdev2000} Much less
is known and understood about the transition driven by the competition
of a strong disorder and interactions in quantum systems which is
the subject of this paper.

The goal of the paper is twofold: to formulate a theoretical model
that is relevant for the description of a number of experimental systems
and to solve this model in the simplest controlled approximation.
The physical systems that we shall focus on are disordered superconductors
but the main results can also be applicable to disordered magnets,
especially disordered ferromagnets in a random field. The new physics
introduced by the strong disorder is the appearance of new phases
\cite{Basko2006} in which all or some excitations are localized in
space and have infinite lifetime and thus cannot contribute to any
transport. The quantum critical point at which the long range order
appears has many features that distinguish it from a conventional
quantum critical point in a translationally invariant systems, most
notably it is characterized by a wide distribution of the order parameter
in a realistic system and the appearance of a new intermediate phase
in which only low energy local excitations have infinitely long lifetime
while high energy excitations can decay.

In Section \ref{sec:Superconductor-insulator-transition:-data} we
analyze the experimental data on superconductor-insulator (SI) transitions
in disordered films of InO, TiN and Be and argue that this transition
is driven by the competition of disorder and superconductivity with
very little effect of the Coulomb repulsion. This will allow us to
formulate a theoretical model for this quantum transition which is
nothing but the model introduced in a seminal work by Ma and Lee \cite{MaLee1985}.
In Section \ref{sec:Formation-of-a-long-range} we develop the formalism
to study the formation of the order parameter in this model at zero
temperatures. In this formalism the appearance of a wide distribution
of the order parameter shows up as replica symmetry breaking. Our
theory provides the justification for the qualitative idea of the
importance of rare sites characterized by a very large susceptibility,
an idea that was first suggested by by Ma, Halperin and Lee \cite{MaHalperinLee}
and observed in the solution of similar one-dimensional models \cite{MaDasgupta,Fisher1992}.
In Section \ref{sec:Width-of-the-levels} we study the properties
of the insulating state. We first determine the level width (decay
rate) at zero temperature and then extend the analysis to low temperatures.
We find two phases in the resulting insulator: an intermediate phase
where only excitations of large enough energy can decay and a third
phase, in the strong disorder regime, characterized by the infinite
lifetime of all excitations, similar to the one proposed in \cite{Basko2006}.
In Section \ref{sec:The-effect-of-frustration} we discuss the effect
of a magnetic field on the phase diagram within our model of the SI
transition. The main conclusion of this Section is that the effects
of the frustration induced by magnetic field are small; this can be
understood as a consequence of the strong inhomogeneity of the order
parameter in the vicinity of the transition. In Section \ref{sec:Consequences-for-experiments.}
we discuss the direct implications of the theory for the experiments
and propose numerical simulations that should test the applicability
of the theory to realistic models. Section \ref{sec:Conclusion} gives
conclusions.

A first quantitative study of the phase diagram of the Ma-Lee model
has appeared recently in \cite{IoffeMezard2010}. The present paper
provides a much more detailed derivation, and studies in detail the
behavior of the order parameter and level width. Similar qualitative
conclusions on the reelvance of the Ma-Lee model and the phase diagram
were reached from phenomenological considerations in the recent paper
\cite{Muller2009}.

\section{Superconductor-insulator transition: data and the model.\label{sec:Superconductor-insulator-transition:-data}}

\subsection{Experimental results\label{sub:Experimenatal-results}}

We begin with the analysis of the data on strongly disordered superconducting
films. Our goal is to formulate the simplest model that captures the
essential physics of these systems. Strongly disordered films of InO,
TiN or Be display a zero field transition from superconductor to insulator
when their resistivity in normal state exceeds a value of the order
of the resistance quantum $R_{Q}=6.5k\Omega$. \cite{SITReview,Gantm2010}
The films in the superconducting state close to the transition become
insulating when subject to a magnetic field.\cite{Hebard1990,Shahar2004,Kapitulnik2008}
The transition driven by magnetic field display a quantum critical
point behavior: the resistance of the films at fields $B<B_{c}$ decreases
with temperature decrease while the resistance of the films at $B>B_{c}$ increases
with temperature decrease. Very important information is provided by tunneling
spectroscopy of such superconducting films in the vicinity of the
quantum critical point\cite{Sacepe2007,Sacepe2008,Sacepe2009,Sacepe2010}.
These data show a well defined gap at all points whilst the coherence
peaks expected for a BCS superconductor appear at some locations and
do not appear at others; a similar phenomenon was reported for high $T_{C}$
oxides\cite{Kapitulnik2001,Davis2001,Davis2002,Yazdani2007}. The
absence of coherence peaks combined with intact superconducting gap
in a single electron tunneling experiment implies that the disorder
does not destroy local Cooper pairing of electrons but prevents formation
of the coherent state of these pairs. This allows to exclude a fermionic
mechanism of superconductivity suppression in these materials: in
such an alternative scenario, the main role of the disorder would
be to enhance the Coulomb interaction that competes with phonon attraction.
This would lead to the suppression of the transition temperature and
the superconducting gap and to the eventual disappearance of both
the superconductivity and the gap. In contrast, the InO and TiN data
demonstrate the presence of a large one-particle gap in the absence
of global coherence and a smooth crossover between superconducting
and insulating gaps as disorder is increased (see \cite{Feigelman2010}
for more detailed discussion).

In the absence of single-electron excitations, the supercondutor-insulator
(SI) transition might happen either because the Coulomb interaction
between the pairs prevents the formation of the condensate, or because
the disorder of Cooper-pair energies prevents their coherent motion
from one site to another. The first scenario might be realized in
the granular materials in which superconductivity remains basically
intact in each grain. The grains are coupled to each other by Josephson
couplings which compete with the Coulomb energy that changes when
a pair moves from one grain to another. Exactly the same physics is
realized in Josephson junction arrays which were extensively studied
some years ago.\cite{vanderZant1996,Fazio2001,Serret2002} In particular,
it was established that in the presence of a magnetic field the Josephson
arrays display a temperature-independent resistance that varies by
many orders of magnitude as a function of a weak magnetic field. In
other words, in these arrays the SI transition does not happen directly,
instead there is a wide region of intermediate 'normal' phase. This
behavior is very surprising given the absence of single electron excitations
in these arrays. It is probably due to the formation of a Cooper-pairs
glass, similar to the electron glass; in this regime collective modes
provide the dissipation mechanism. Although the theoretical picture
of this phase is not clear, the observed experimental behavior of
these systems is in a striking contrast with the behavior of disordered
films which show a direct transition between superconductor and insulator.

This leaves the only possible mechanism for the superconductor-insulator
transition in disordered films: the competition between pair hopping
and random pair energies on different sites, as suggested 25 years
ago in a seminal paper of Ma and Lee \cite{MaLee1985} (see also
\cite{BulSad,KotKap,Randeria}).
As we show in the present paper, the solution of this model reproduces
correctly the most important features of the data: direct SI transition,
activated behavior close to the quantum critical point in the insulating
phase, strong dependence of the activation energy near the quantum
critical point and huge order parameter variations from site to site
in the superconducting phase.

Recent experimental data indicate the possibility of a faster than
activated temperature dependence of the resistance in the insulating
phase (characterized by growing $T\ln R$ at low $T$). This behavior
is very unusual for disordered electron systems which typically display
either activated, Efros-Shklovskii or Mott behavior characterized
by $\ln R\propto T^{-\alpha}$ with $\alpha\leq1$. It can be understood
if the pair excitations remain localized in space at low energies
but are delocalized at high energies with the temperature dependent
mobility edge that separates them.

For the microscopic justification of this mechanism one needs to find
the reason why Coulomb repulsion does not play a role in the 
superconductor - insulator transition in some disordered films. Phenomenologically,
it is known~\cite{Zvi} that the dielectric constant in these materials
remains very large, $\kappa\geq30$ deep in the insulating phase of
InOx. A large value of the dielectric constant between low energy
electrons close to the Fermi surface allows one to neglect the effect of Coulomb 
interaction 
on the pairing even for relatively strong disorder that results in
wave functions localization. Because the host matrix dielectric constant
can only increase with additional carriers, the Coulomb interaction
between conducting electrons close to the Fermi surface is strongly
suppressed in this materials. The microscopic reason for that might
be a very strong energy dependence of the density of states (in particular,
it was computed for a typical small cluster of TiN\cite{Anisimov}),
this implies that although density of states exactly at the Fermi
level is small and the states there get localized, there are many
single electron states within $10meV$ range that provide large screening
of the bare Coulomb interaction. 

A detailed analysis of the competition between Cooper pairing and
wave-function localization is provided by a recent paper\cite{Feigelman2010}.
In particular, it shows that global superconductivity can survive
in a range of relatively strong disorder. Within this scenario the
localized electron pairs are formed at high temperatures when two
electrons bind together (with typical binding energy $\Delta_{P}$)
on a weakly localized state. At lower temperatures, $T_{c}\leq\Delta_{P}$,
coherent hopping of these pairs leads to the long-range coherence
and to the formation of a superconducting condensate. Direct experimental
confirmation of this scenario is provided by the scanning tunneling
microscopy data\cite{Sacepe2007,Sacepe2008,Sacepe2009,Sacepe2010}.
At even stronger disorder long-range coherence is never formed, the
resulting state is insulating, characterized by a large single-particle
gap. 

The paper \cite{Feigelman2010} studied the properties of the superconducting
and insulating phases far from the transition. In this work we study
the properties of these phases in the vicinity of the transition in
which the characteristic energy scales are much smaller than the single
particle gap, $T_{c}\ll\Delta_{P}$. In this study we shall focus
on these low energy scales and neglect the contribution from single-electron
occupied localized orbitals. This allows us to describe the physics
entirely in terms of Anderson pseudospins~\cite{Anderson1959}. The
interaction between the pseudospins is due to the matrix elements
of the Cooper attraction between original electrons. As discussed
in the work \cite{Feigelman2010} these matrix elements are modified by
the fractal nature of the electron wave functions which leads to correlations
and to a smooth dependence of $T_{c}$ on disorder strength. In the present
paper we shall ignore the effect of fractality  that should not affect the qualitative
properties in the vicinity of SI transition. In the absence of such
correlations the \emph{average} interaction between pseudospins does
not depend on disorder. Thus, when comparing with experimental data
we shall assume that changing of the disorder translates \emph{only}
into  changing  the number of "interacting neighbours" of a given pseudospin.

\subsection{The model: localized pairs and pseudospin representation. \label{sub:The-model:-localized}}

In the absence of Coulomb repulsion the only interaction between electrons
is the attraction that leads to their pairing on the localized single
electron states.\cite{Feigelman2010} The strength of this pairing
is inversely proportional to the volume of the state. Because of the
fractal nature of the single electron wave functions the volume of
a typical localized state is small, this makes the pairing energy
large. Thus, the low energy degrees of freedom in this system are
pairs that can hop from one site to another. This physics is described
by a Hamiltonian of disordered hard-core bosons, or, equivalently,
pseudospin operators (originally introduced by Anderson\cite{Anderson1959}
for the pure system, and later by Ma and Lee\cite{MaLee1985} for
the disordered system): \begin{equation}
H_{XY}=-2\sum_{i}\xi_{i}s_{i}^{z}-\sum_{(ij)}M_{ij}(s_{i}^{+}s_{j}^{-}+s_{i}^{-}s_{j}^{+})\label{H_A}\end{equation}
 where $\mathbf{s}_{i}=\frac{1}{2}\mathbf{\sigma}_{i}$ are spin-$\frac{1}{2}$
operators and the sum $\sum_{(ij)}$ goes over all different pairs
of neighbors $i,j$. The state with $s_{i}^{z}=\pm\frac{1}{2}$ corresponds
to a local level occupied or unoccupied by a Cooper pair; $\xi_{j}$'s
are occupation energies for each site, which are quenched random variables
drawn from a probability distribution $p(\xi)$. Hereafter we assume
a box distribution $p(\xi)=\frac{1}{2W}\theta(W-|\xi|)$, so that
the non-interacting density of states is $\nu=1/2W$. The important
feature of this distribution is that it is constant near to $\xi=0$;
the value of $\nu$ just sets the scale of energies, and we shall
choose $\nu=1$. The matrix elements $M_{ij}$ describe the hopping
amplitudes of Cooper pairs. These hopping amplitudes couple a typical
local level to a large number of neighbors, $Z\gg1$. We shall assume
that each site is coupled to $Z$ neighbors with $M_{ij}=2g/(Z-1)$.
Another closely related problem corresponds to the Ising ferromagnet
in a random transverse field; the model is defined by the Hamiltonian
\begin{equation}
H_{I}=-\sum_{i}\xi_{i}\sigma_{i}^{z}-\frac{g}{Z-1}\sum_{(ij)}\sigma_{i}^{x}\sigma_{j}^{x}\label{H_B}\end{equation}
 For brevity we shall refer to this model as Ising model below. We
shall mainly study the XY problem (\ref{H_A}) but most of our conclusions
also hold for the case (\ref{H_B}). As will be clear below, in the
leading approximation the two models (\ref{H_A}) and (\ref{H_B})
lead to identical cavity equations for both the order parameter and
the relaxation rate.

The Hamiltonian (\ref{H_A}) describes the superconductor-insulating
transition as a ferromagnetic spin $\frac{1}{2}$ system with random
transverse fields, as proposed originally in\cite{Anderson1959,MaLee1985}.
In this language the superconducting phase corresponds to the existence
of a spontaneous magnetization in the $x-y$ plane.

\section{Formation of a long-range superconducting order. \label{sec:Formation-of-a-long-range}}

\subsection{A simple mean-field approach \label{sub:A-simple-mean-field}}

The most obvious approach to study the Hamiltonians (\ref{H_A},\ref{H_B})
is to use the mean-field approach, in which $H_{XY}$ is replaced
by $H_{MF}=\sum_{i}(-\xi_{i}\sigma_{i}^{z}-B\sigma_{i}^{x})$ where
$B$ is determined self-consistently from the equation $B=(g/(Z-1))\sum_{j}\langle\sigma_{j}^{x}\rangle$.
The right-hand side of this equation contains a sum of $Z\gg1$ terms,
so it might be expected to become site-independent in the large $Z$
limit. Although we shall challenge below the assumptions and conclusions
of this approach, it is instructive to summarize its main results.
The mean field Hamiltonian on site $i$ has two eigenstates with energies
$-\tau_{i}\sqrt{\xi_{i}^{2}+B^{2}}$ where $\tau_{i}=\pm1$. At temperature
$T=1/\beta$, one obtains the self consistent equation for $B$, which
gives either $B=0$ or \begin{equation}
1=g\int d\xi\, p(\xi)\,\frac{\tanh(\beta\sqrt{\xi^{2}+B^{2}})}{\sqrt{\xi^{2}+B^{2}}}\ .\label{eq:BCS_equation}\end{equation}
 At zero temperature this equation has a solution for any $g$, so
the ground state is characterized by a non-zero field, $B>0$. This
corresponds to a spontaneous order parameter (in spin language, this
is a non-zero magnetization) in the $x$ direction, so the system
is a superconductor (or a ferromagnet in case of the Ising model).
At finite temperature there is a transition from an insulating to
a superconducting phase, when (\ref{eq:BCS_equation}) starts to have
a solution. Assuming a constant density of states, $p(\xi)=\frac{1}{2}\theta(1-|\xi|)$,
we find that the critical temperature is determined by: \begin{equation}
\frac{1}{g}=\frac{1}{2}\int_{-1}^{1}d\xi\frac{\tanh(\xi/T_{c})}{\xi}\quad\Longrightarrow\quad T_{c0}=\frac{4e^{\mathbf{C}}}{\pi}\exp(-1/g)\label{gc_naive}\end{equation}
 where $\mathbf{C}\approx0.577..$ is the Euler number. As we show
below, while this mean field prediction is correct at $Z=\infty$,
it is qualitatively wrong for a system with a finite connectivity
(even if $Z$ is large) in the limit of very small $g$.

\subsection{Cavity mapping\label{sub:Cavity-mapping}}

We now turn to a more refined mean-field discussion, valid for finite
$Z\gg1$, which is the basis for our main results. We shall employ
the cavity method that has been developed in the classical case to
study frustrated systems on a Bethe lattice, a fixed connectivity
random graph of connectivity $Z$, see \cite{MezardParisi2001}. Its
generalization to quantum problems without disorder has been studied
in \cite{Laumann,Krzakala}, using the Suzuki-Trotter formalism. We
shall use here an approximation of the quantum cavity method which
does not use the Suzuki-Trotter formalism, and allows to study analytically
the quantum disordered systems. As we shall see below (see discussion
below (\ref{eq:K_RSB}) and section \ref{sub:Leading-correction-in}),
this method takes into account the leading corrections to the naive
mean field at large-but-finite $Z$, which turn out to be of order
$1/\ln Z$ . It neglects the corrections that are small in $1/Z$.
Appendix A explains the relation of our approach to the full Suzuki-Trotter
quantum cavity method.

In the simplified cavity method one emulates the effects of spin environment
by the static field acting on it. One thus studies the properties
of a spin $j$ in the cavity graph where one of its neighbors has
been deleted, assuming that the $K=Z-1$ remaining neighbors are uncorrelated.
The system of spin $j$ and its $K$ neighbors is thus described by
the local Hamiltonian

\begin{equation}
H_{j}^{cav}=-\xi_{j}\sigma_{j}^{z}-\sum_{k=1}^{K}\left(\xi_{k}\sigma_{k}^{z}+B_{k}\sigma_{k}^{x}+\frac{g}{K}(\sigma_{j}^{x}\sigma_{k}^{x}+\sigma_{j}^{y}\sigma_{k}^{y})\right)\label{eq:H^cav}\end{equation}
 where $B_{k}$ is the local {}``cavity'' field on spin $k$ due
to the rest of the spins (in absence of $j$). Here we have chosen
the $x$ direction as the transverse direction where the spontaneous
ordering takes place. By solving the problem of $Z$ spins in (\ref{eq:H^cav}),
one can compute the induced magnetization of $j$, $\langle\sigma_{j}^{x}\rangle$,
which is by definition equal to $\frac{B_{j}}{\sqrt{\xi_{j}^{2}+B_{j}^{2}}}\tanh\beta\sqrt{\xi_{j}^{2}+B_{j}^{2}}$.
One thus gets a mapping allowing to compute the new cavity field $B_{j}$
in terms of the $K$ fields $B_{k}$ on the neighboring spins. This
mapping induces a self-consistent equation for the distribution of
the $B$ fields \cite{MezardParisi2001}. Notice that the use of the
mean-field approximation in order to study the cavity mapping is very
different from its naive use in (\ref{H_B}).

One can make one more approximation which simplifies the cavity mapping
and makes possible an analytical study. Similarly to the mean-field
approach discussed above, one can replace dynamical spin variables
$\sigma_{k}^{x}$ by their averages. This approximation neglects some
terms of order of $1/K$; below we refer to it as the cavity-mean-field
approximation. We shall first study this approximation here. Later
on, in section \ref{sub:Leading-correction-in}, we shall compare
its results with the ones found by the full cavity mapping. As we
show in this section the cavity-mean-field approximation becomes very
accurate even for moderately small $K$. The physical reason for this
is that the main effect missed by the cavity-mean-field approximation
is the level repulsion induced by the interaction between sites $j$
and $k$. However this repulsion happens only in the rare cases when
$\xi_{k}$ is close to $\xi_{j}$, and it has a significant effect
on the resulting value of the local order parameter only when both
of them are close to $0$. Because this happens very rarely, the cavity-mean-field
approximation turns out to be very good for this problem.

Formally, the cavity mean field approximation amounts to approximating
the cavity Hamiltonian (\ref{eq:H^cav}) acting on spin $j$ by \begin{equation}
H_{j}^{cav-MF}=-\xi_{j}\sigma_{j}^{z}-\sigma_{j}^{x}\frac{g}{K}\sum_{k=1}^{K}\langle\sigma_{k}^{x}\rangle\label{eq:H^cav-MF}\end{equation}
 This implies that $B_{j}=\frac{g}{K}\sum_{k=1}^{K}\langle\sigma_{k}^{x}\rangle$,
giving the recursion equation relating the $B$ fields: \begin{equation}
B_{j}=\frac{g}{K}\sum_{k=1}^{K}\frac{B_{k}}{\sqrt{B_{k}^{2}+\xi_{k}^{2}}}\tanh\beta\sqrt{B_{k}^{2}+\xi_{k}^{2}}\ .\label{eq:mapping_Kfinite}\end{equation}

This recursion induces a self-consistent equation for the distribution
$P(B)$: \begin{equation}
P(B)=\int dB_{1}\dots dB_{K}P(B_{1})\dots P(B_{K})\delta\left(B-\frac{g}{K}\sum_{k=1}^{K}\frac{B_{k}}{\sqrt{B_{k}^{2}+\xi_{k}^{2}}}\tanh\beta\sqrt{B_{k}^{2}+\xi_{k}^{2}}\right)\ .\label{eq:PdeB_Kfinite}\end{equation}
 This equation always allows the trivial solution $P(B)=\delta(B)$
corresponding to an insulator. The superconducting transition is signaled
by the appearance of another solution characterized by a non-trivial
distribution function $P(B)$. It turns out that this transition and
the properties of the phases in its vicinity can be best studied using
methods developed in the statistical physics of random systems.

\subsection{The directed polymer problem and replica symmetry breaking\label{sub:The-directed-polymer}}

\begin{figure}[ht]
 \includegraphics[width=7cm]{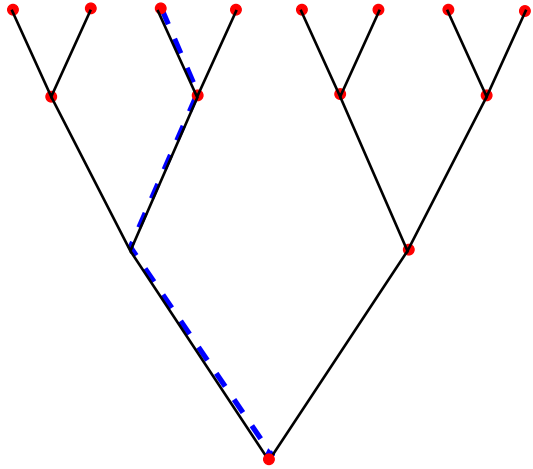} \caption{Iteration of the cavity equations three times (here with $K=2$).
The field $B_{0}$ on the root is proportional to the infinitesimal
field $B$ added on the boundary. The corresponding susceptibility,
given in (\ref{eq:Xi_def}), is equal to the partition function of
the DP problem, obtained by summing over all the paths from the root
to the boundary (such as the one shown by the dashed line). In this
sum, each path has a weight equal to the product of terms associated
with each edge that it visits.}

\label{fig:DP_Tree} 
\end{figure}

In order to understand the properties of the mapping (\ref{eq:mapping_Kfinite}),
let us imagine that we iterate it $L\gg1$ times on a Bethe lattice.
For $L$ finite and $N\rightarrow\infty$ the corresponding graph
is just a rooted tree with branching factor $K$ at each node and
depth $L$ (see Fig. \ref{fig:DP_Tree}). The field $B_{0}$ at the
root is a function of the $K^{L}$ fields on the boundary. In order
to see whether there is spontaneous ordering, we study the value of
$B_{0}$ in linear response to infinitesimal fields $B_{i}=B\ll1$
on the boundary spins. This is given by \begin{equation}
B_{0}/B=\Xi\equiv\sum_{P}\prod_{k\in P}\left[\frac{g}{K}\frac{\tanh(\beta\xi_{k})}{\xi_{k}}\right]\ .\label{eq:Xi_def}\end{equation}
 where the sum is over all paths going from the root to the boundary,
and the product $\prod_{n\in P}$ is over all edges along the path
$P$. The response $\Xi$ is nothing but the partition function for
a directed polymer (DP) on a tree, where the energy of each edge is
$e^{-E_{k}}=(g/K)(\tanh(\beta\xi_{k})/\xi_{k})$ and the temperature
has been set equal to one. The general method for the solution of
such problems has been developed by Derrida and Spohn\cite{DerrSpohn}.
The solution can be expressed in terms of the convex function \begin{equation}
f(x)=\frac{1}{x}\ln\left[K\int_{-1}^{1}\frac{d\xi}{2}\;\left(\frac{\tanh(\beta\xi)}{\xi}\right)^{x}\right]\ .\label{eq:f(x)}\end{equation}
 Let us denote by $x=m$ the value of $x$ where this function is
minimal. In the large $L$ limit, there exist two regimes for the
DP problem: 
\begin{itemize}
\item {}``RS-phase'': If $m>1$, then $(1/L)\ln\Xi=f(1)+\ln(g/K)$. In
this case the superconducting phase appears at a value $g_{c}=Ke^{-f(1)}$
which is the same result as found by the naive mean field approach
in (\ref{gc_naive}). 
\item {}``RSB-phase'': If $m<1$, then $(1/L)\ln\Xi=f(m)+\ln(g/K)$. The
ordered phase appears at $g_{c}=Ke^{-f(m)}$ which is larger than
the naive estimate (\ref{gc_naive}). 
\end{itemize}
The method for obtaining these results developed by Derrida and Spohn\cite{DerrSpohn}
employed a mapping to the traveling wave equations. An alternative
way, which we now summarize, is to use the replica method, similar
to the one used in\cite{CookDerrida90}. It gives a compact solution
and helps to understand the physical meaning of the DP phase transition.
The names of the two phases used above ({}``RS'' stands for replica
symmetric'' and {}``RSB'' stands for replica-symmetry broken'')
originate from this analysis.

The DP partition function $\Xi$, defined in (\ref{eq:Xi_def}), depends
on the random quenched variables $\xi_{n}$. One expects that the
free-energy in a short-range-interaction problem is a self-averaging
quantity, so the value of $\log\Xi$ for a \emph{typical} sample is
obtained from the quenched average of $\log\Xi$ over these random
variables, denoted by $\overline{\ln\Xi}$ . In the replica method
one computes it by writing \begin{eqnarray}
\overline{\ln\Xi} & = & \lim_{n\rightarrow0}(\overline{\Xi^{n}}-1)\nonumber \\
\Xi^{n} & = & \prod_{a=1..n}\left(\sum_{P_{a}}\prod_{k\in P_{a}}\frac{g}{K}\frac{\tanh(\beta\xi_{k})}{\xi_{k}}\right)\ .\end{eqnarray}
 The average of $\Xi^{n}$ is obtained by a sum over $n$ paths, \begin{eqnarray}
\overline{\Xi}^{n}=\sum_{P_{1},\dots,P_{n}}\prod_{k}\overline{\left(\frac{g}{K}\frac{\tanh(\beta\xi_{k})}{\xi_{k}}\right)^{r_{k}}}\ ,\label{xinav}\end{eqnarray}
 where the weight of each edge $k$ in the tree depends on the number
$r_{k}$ of paths which go through this edge.

The RS solution assumes that the leading contribution to (\ref{xinav})
comes from non-overlapping independent paths ($r_{k}=1$). This gives
\begin{equation}
\overline{\Xi^{n}}=K^{Ln}\left[\frac{g}{K}\int_{-1}^{1}\frac{d\xi}{\xi}\tanh(\beta\xi)\right]^{Ln}=\exp(Ln[\log(g/K)+f(1)])=\left(\overline{\Xi}\right)^{n}\ .\end{equation}

The RSB solution assumes that the leading contribution to (\ref{xinav})
comes from patterns of $n$ paths which consist of $n/x$ groups of
$x$ identical paths, where the various groups go through distinct
edges. This gives: \begin{equation}
\overline{\log\Xi}=\frac{L}{x}\;\ln K\ \overline{\left(\frac{g}{K}\frac{\tanh(\beta\xi_{k})}{\xi_{k}}\right)^{x}}\equiv L\left[\ln\left(\frac{g}{K}\right)+f(x)\right]\end{equation}
 where $f(x)$ is the function introduced in (\ref{eq:f(x)}). In
the replica limit $n\rightarrow0$, the parameter $x$ should belong
to the interval $[0,1]$. Minimizing the function $f(x)$ over $x\in[0,1]$
(the fact that one should minimize $f$, and not maximize it, is a
well-known aspect of the replica method \cite{MezardParisiVirasoro}),
one gets the phase diagram described above: there exists a critical
value of the inverse temperature $\beta=\beta_{RSB}$ such that, for
$\beta<\beta_{RSB}$, the function $f(x)$ is minimal at the boundary,
$x=1,$ of the interval $[0,1]$; the DP is in its replica symmetric
phase. For $\beta>\beta_{RSB}$, the function $f(x)$ has a minimum
inside the interval $(0,1)$ at some value $x=m<1$, this corresponds
to the spontaneous breakdown of the replica symmetry in the DP problem.

These two regimes of the DP problem are qualitatively very different.
In the RS regime the measure on paths defined in (\ref{eq:Xi_def})
is more or less evenly distributed among all paths. On the contrary,
the RSB regime is a glass phase where the measure condenses onto a
small number of paths. An order parameter which distinguishes between
these two phases is the participation ratio $Y=\sum_{P}w_{P}^{2}$,
where $w_{P}$ is the relative weight of path $P$ in the measure
(\ref{eq:Xi_def}). It is easy to see that $Y=0$ in the RS phase.
In the RSB phase, the value of $Y$ is finite and non self-averaging
(it depends on the realization of the $\xi$'s), and its average is
given by $1-m$. This glass transition, and the nature of the RSB
glass phase, are identical to the ones found in the random energy
model \cite{DerridaREM,GrossMezard}.

\subsection{Phase diagram\label{sub:Phase-diagram}}

The results on the DP problem described in the previous section allow
to derive the phase diagram of disordered superconductors. The state
of the model is determined by the three parameters: the coupling $g$,
the cavity connectivity $K=Z-1$, and the temperature $T=1/\beta$.
The phase transition between insulator and superconductor is a surface
in this three-dimensional parameter space. Depending on the purpose,
it can be useful to view it in different directions. We shall define
the phase transition as $g_{c}(K,T)$, or $K_{c}(g,T)$, or $T_{c}(K,g)=1/\beta_{c}(K,g)$.
In the zero-temperature limit we shall speak of $g_{c}(K)=g_{c}(K,T=0)$
and $K_{c}(g)=K_{c}(g,T=0)$.

The DP partition function $\Xi$ gives the susceptibility, measured
on the root site, to a small field on the boundary. When this susceptibility
diverges, there is spontaneous generation of non-zero $B$ fields,
i.e. the systems is in the superconducting phase. The superconducting
phase transition point is given by: 
\begin{itemize}
\item If $\beta<\beta_{RSB}$ (RS phase of the DP), then $g_{c}=Ke^{-f(1)}$.
This coincides with the result of the naive mean field analysis. 
\item If $\beta>\beta_{RSB}$ (RSB phase of the DP), the phase transition
is obtained by solving the two equations $f(m)+\ln\frac{g_{c}}{K}=0$
and $f^{\prime}(m)=0$. 
\end{itemize}
At zero temperature the function $f(x)$ can be computed analytically:\begin{equation}
f(x)=\frac{1}{x}\ln\left(\frac{K}{1-x}\right)\label{eq:f0}\end{equation}
 leading to two equations for $m$ and $g_{c}$ which have the explicit
solution \begin{eqnarray}
g_{c}e^{1/(eg_{c})} & = & K\label{eq:Kc}\\
m & = & 1-eg_{c}\label{eq:xsol}\end{eqnarray}
 The value of $g_{c}$ determined by the solution of (\ref{eq:Kc})
gives the location of the zero temperature quantum phase transition
$g_{c}(K)$ between the insulating and the superconducting phase.
Note that in the framework of the recursion equations (\ref{eq:mapping_Kfinite})
this is an exact result. It predicts a finite value of $g_{c}$ for
all $K\ge1$. In particular, when $K=1$ it reproduces correctly the
result $g_{c}=1/e$ known for the one-dimensional chain \cite{Fisher1992}.
This is in contrast with the naive mean field equation (\ref{gc_naive})
which wrongly predicts $g_{c}=0$ for all $K$. For the comparison
with the experiment it will be useful to describe the phase transition
point as a function of the effective number of cavity neighbors $K$
for a given a value $g$ of the hopping (see section \ref{sub:Experimenatal-results});
in terms of this parameter the system becomes superconducting at $K>K_{c}(g,T)$.
The zero temperature transition point is given by explicit equation:
\begin{equation}
K_{c}(g)=ge^{1/(eg)},\ \ \ g\le1/e.\label{eq:K_c}\end{equation}

We now discuss the finite temperature phase transition. At sufficiently
high temperatures, $T>T_{RSB}=1/\beta_{RSB}$, the transition line
$T_{c}(K,g)$ is independent of $K$ and is given by the naive mean
field result (\ref{gc_naive}): $T_{c}=(4e^{C}/\pi)e^{-1/g}$. The
RSB solution appears at a value $g_{RSB}$ such that $T_{c}(K,g_{RSB})=T_{RSB}$.
At this $g$ the derivative $f'(1)=0$, which gives two equations
that determine the position of this point on $(g,T)$ or $(K,T)$
plane: \begin{eqnarray}
g\int_{0}^{1}\frac{d\xi}{\xi}\tanh\beta\xi & = & 1\\
g\int_{0}^{1}\frac{d\xi}{\xi}\ln\left(\frac{g}{\xi K}\tanh\beta\xi\right)\tanh\beta\xi & = & 0\label{eq:Eq_K_RSB}\end{eqnarray}
 The result is again best expressed by inverting $g_{RSB}(K)$ in
order to get the critical value $K_{RSB}(g)$ below which the naive
mean field solution breaks down. The asymptotic of the solution of
the equations at $g\ll1,\ K\gg1$ is \begin{equation}
K^{RSB}(g)=ge^{1/(2g)}\label{eq:K_RSB}\end{equation}
 The equations for the critical values of the number of neighbors
(\ref{eq:K_c},\ref{eq:K_RSB}) show that, at large $K$, the solution
deviates from the naive mean field solution of section \ref{sub:A-simple-mean-field}
at $g\lesssim1/(e\ln K)$. This demonstrates that our approximation
takes into account $1/\ln K$ corrections, as announced above. Note
that at weak coupling $K^{RSB}(g)$ is exponentially larger than $K_{c}(g)$,
so that there is a broad range of $K$ where RSB effects are important.

An important unexpected qualitative conclusion from (\ref{eq:K_RSB})
is that the \textit{effective} number of neighbors of a given site,
$Z_{\mathrm{eff}}$, defined as the number of neighbors having local
energies $\xi_{j}$ within the energy stripe $\pm T_{c}$, is typically
much smaller than unity. Indeed, at the temperature $T_{RSB}$, this
number is given by \begin{equation}
Z_{\mathrm{eff}}=K^{RSB}T_{c0}=\frac{4}{\pi}e^{\mathrm{C}}ge^{-1/2g}\ll1,\label{Zeff}\end{equation}
 it decreases with $K$ roughly as $Z_{\mathrm{eff}}\propto1/K$ when
$K\gg1$. Although $Z_{eff}$ is small, it leads to a non-zero $T_{c}$
and to RSB effects. This is another signature of the fact that the
transition is governed by rare sites.

\begin{figure}[ht]
\includegraphics[width=4in]{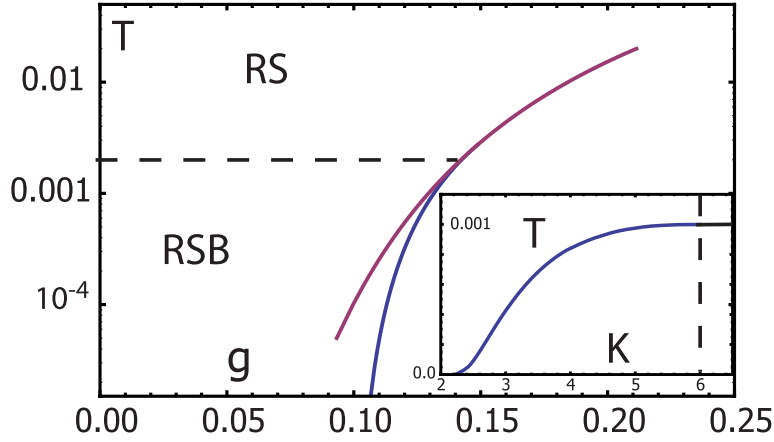} \caption{Main panel: phase diagram in plane $(g,T)$ for $K=4.$ The full lines
show the critical temperature as function of $g$. The low temperature
phase is superconducting, the high temperature phase is insulating.
The top curve is the naive mean-field prediction which gives the correct
result above $T_{RSB}=0.0207$. The bottom curve is the result of
the analysis on the Bethe lattice described in the text that includes
the RSB effects in the DP problem, which occur at temperatures $T<T_{RSB}$.
The insert shows the critical temperature as function of $K$ for
$g=0.129$, a value which roughly corresponds to the experimental
situation in disordered InO films (see section \ref{sec:Consequences-for-experiments.}).
For this value of $g$, the replica symmetric solution gives a $K$-independent
transition temperature $T_{c}=0.001$. This prediction of the replica
symmetric theory is correct for $K>K^{RSB}\simeq6$. For smaller $K$
the transition temperature starts to drop, the quantum critical point
corresponds to $K_{c}\simeq2.2$. Notice that in a rather wide regime
the replica symmetry is broken but the effect of the breaking on the
transition temperature is small. }

\label{fig:PhaseDiagram} 
\end{figure}

The phase diagram for $K=4$ is shown in Fig. \ref{fig:PhaseDiagram}.
At any temperature, there is a finite critical value of the coupling,
$g_{c}(K,T)$, separating an {}``ordered\textquotedblright{}, superconducting,
phase with spontaneous order parameter at $g>g_{c}(K,T)$ from a 
{}``disordered\textquotedblright{}  normal phase with zero order
parameter at $g<g_{c}(K,T)$. Contrary to the naive MF prediction,
$g_{c}(K,T)$ remains non-zero at $T=0$.

One of the important characteristic properties of the quantum critical
point is the behavior of $T_{c}(K,g)$ near to the critical point
$K_{c}(g)=ge^{1/(eg)}$ where it drops to zero. In order to study
this behavior, we look for the solution of the two equations $f(x)+\log(g/K)=0$
and $f'(x)=0$. The explicit form of these equations is \begin{eqnarray}
K\left(\frac{g}{K}\right)^{x}\int_{0}^{1}d\xi\frac{\tanh^{x}\beta\xi}{\xi^{x}} & = & 1\label{eq:Tc_nearKc1}\\
\int_{0}^{1}d\xi\frac{\tanh^{x}\beta\xi}{\xi^{x}}\ln\left(\frac{g}{\xi K}\tanh\beta\xi\right) & = & 0\label{eq:Tc_nearKc2}\end{eqnarray}
 We assume that $K=K_{c}(g)(1+\delta)$ with $\delta\ll1$, and we
introduce the quantity $y=1-x$, which is expected to be close to
its critical value $y_{c}=eg$. The second equation (\ref{eq:Tc_nearKc2})
leads to the relation $y(\delta)=y_{c}(1-y_{c}\delta)$. Next, in
order to determine $T_{c}(\delta)$ we use the stationarity of the
first equation with respect to the simultaneous variations of $T$,
$K$ and $y$, and find \begin{equation}
T_{c}(K)=\vartheta(y_{c})\left(\frac{K}{K_{c}}-1\right)^{1/y_{c}}\,;\qquad\vartheta(y)=\left[\frac{y}{(1-y)C(y)}\right]^{1/y}\label{eq:Tc(K)}\end{equation}
 with \[
C(y)=\int_{0}^{\infty}dt\left(\frac{t}{\tanh t}\right)^{y}\cosh^{-2}t\,\]
 The result (\ref{eq:Tc(K)}) is valid as long as $T_{c}(K)$ is much
less than the mean-field transition temperature $T_{c0}\approx e^{-1/g}$
and $K-K_{c}\ll K_{c}$. In terms of $K-K_{c}$ the former condition
is \begin{equation}
\left\vert \frac{K}{K_{c}}-1\right\vert \ll\frac{1}{e^{1+e}g}\approx\frac{1}{50g}\label{KKc}\end{equation}
 For physically relevant systems $g\gg0.02$, so the condition (\ref{KKc})
determines the regime of $K$ for which the result (\ref{eq:Tc(K)})
holds. This analysis shows that the transition temperature $T_{c}(K)$
goes down very slowly when $K$ decrease below the $K^{RSB}(g)$ value,
until it reaches the close vicinity of $K_{c}\ll K^{RSB}$ where it
drops sharply to zero according to Eq.(\ref{eq:Tc(K)}). The numerical
solution of Eqs.(\ref{eq:Tc_nearKc1},\ref{eq:Tc_nearKc2}) for $T_{c}(K,g)$
at $g=0.129$ is presented in Fig.\ref{fig:PhaseDiagram} (insert).

The replica symmetry breaking at $K<K^{RSB}$, which is at the origin
of the failure of the naive mean field analysis, has important physical
consequences. Physically it signifies the absence of self averaging
which is due to the importance of very rare sites with small $\xi_{j}$
that easily polarize in the $x$-direction, becoming thus nucleation
centers for the ordered phase. The qualitative importance of these
rare sites and resulting absence of self averaging was first discussed
in \cite{MaHalperinLee}. Our analysis gives the mathematical formalism
to evaluate these effects and their consequences within the well-controlled
Bethe approximation, such as the exact position of the phase transition
point. The role of rare sites with small $\xi_{j}$ can be quantitatively
characterized by the condensation of the measure in the DP problem.
Let us imagine adding a small field in the $x$ direction on one site
$j$, and let us look at its effect on the spins $k$ at distance
$L$ from $j$; this gives the picture of the correlations in the
many-body wave-function. In the RSB phase, the order develops only
along a small number of paths; this fact is completely missed by the
naive mean field approach.

Even at $K>K^{RSB}$ the self averaging is not fully restored. In
particular, in a wide range of $K$ the simple mean-field approximation
is not applicable for the calculation of higher moments of the $P(B)$
distribution. The upper border of this region is given by the value
$K^{\ast}$ corresponding to the divergence of infinitely high moments
(divergence of $\langle B^{n}\rangle$ at $n\rightarrow\infty$).
The condition for this divergence is given by the inequality $g/KT_{c}\geq1$.
With $T_{c}$ determined by the mean-field approximation, we get the
critical value $K^{\ast}$ that corresponds to the divergence of large
moments: \begin{equation}
g\int_{0}^{K^{\ast}/g}\frac{dz}{z}\tanh z=1\,\quad\Longrightarrow K^{\ast}=\frac{\pi}{4}e^{-\mathrm{C}}ge^{1/g}\label{Kstar}\end{equation}
 where $\mathrm{C}$ is the Euler number. The full self-averaging
is expected only at $K>K^{*}$.

\subsection{Distribution function of the local order parameters and the phase
diagram. \label{sub:Distribution-function-of}}

In this section we derive the equation for the distribution function
$P(B)$ of local fields close to the transition point, $T\approx T_{c}(g,K)$,
using the cavity mapping (\ref{eq:mapping_Kfinite}) as the starting
point. We will use the Laplace transform of this distribution, $\mathcal{P}(s)=\int_{0}^{\infty}dBP(B)e^{-sB}$.

We start from the linearized version of Eq.(\ref{eq:mapping_Kfinite}),
\begin{equation}
B_{i}=\frac{g}{K}\sum_{k}\frac{B_{k}}{\xi_{k}}\tanh(\beta\xi_{k})\ ,\end{equation}
 which is adequate to describe the phase transition region where the
fields are small. The Laplace transform satisfies the equation: \begin{equation}
\mathcal{P}(s)=\left[\int_{0}^{1}d\xi\,\,\mathcal{P}\left(s\frac{g}{K}\frac{\tanh\beta\xi}{\xi}\right)\right]^{K}\label{eq:Laplace_recursion}\end{equation}
 Consider the first terms of its expansion at small $s$, and assume
that this expansion behaves as $\mathcal{P}(s)=1-As^{x}$. If $x=1$,
the mean $\int dBP(B)B$ is finite. We will see that this is the situation
in the RS phase. The situation $x<1$ occurs in the RSB phase, and
corresponds to a distribution $P(B)$ which decays at large $B$ like
$B^{-1-x}$, with a diverging mean.

Let us first assume that $x=1$. Then $A$ is just the mean of $B$,
i.e. it coincides with the usual superconducting order parameter.
Plugging $\mathcal{P}(s)=1-As$ into (\ref{eq:Laplace_recursion}),
the resolvability condition for the linear equation in $A$ leads
immediately to the standard mean-field equation for $T_{c}$, see
(\ref{gc_naive}). We learned previously from the replica analysis
that this equation is valid only in the RS regime: it is not correct
when $K<K_{c}(g)$. Therefore, RSB corresponds to the divergence of
the mean order parameter. To study the RSB phase, we assume that $\mathcal{P}(s)=1-As^{x}$
with $x<1$. From (\ref{eq:Laplace_recursion}), we obtain the condition
\begin{equation}
1=K\int\frac{d\xi}{\xi}\left(\frac{g}{K}\frac{\tanh(\beta\xi)}{\xi}\right)^{x}\end{equation}
 which can also be written by using the function $f$ introduced in
the replica analysis as $f(x)+\log(g/K)=0$. Therefore, for any $x<1$,
there is a non-trivial solution $g=Ke^{-f(x)}$. In this situation,
it is reasonable to assume that the critical value of $g$ is the
largest one among all these values, which means that it is obtained
by finding the minimum of $f(x)$. We shall prove this assumption
in section \ref{sub:Spatial-scale-of} where we study the spatial
evolution of the distribution function. This is precisely the result
obtained with the replica solution in section \ref{sub:Phase-diagram}
of the DP problem, with $x=m$ defined in Eq.(\ref{eq:xsol}), a solution
which coincides with the result  obtained by travelling waves
analysis \cite{DerrSpohn} when applied to this problem. The corresponding
distribution function $P(B)$ is characterized by a power-law behavior
at large $B\gg B_{0}$, where $B_{0}$ is of the order of a typical
value of $B$: \begin{equation}
P(B)=\frac{B_{0}^{m}}{B^{1+m}}\label{eq:P(B)_scaling_form}\end{equation}
 This power-law tail at large $B$ translates into the behavior of
the Laplace transform: \begin{equation}
\mathcal{P}(s)\approx1-(sB_{0})^{m}\quad\mathrm{at}\quad sB_{0}\ll1\label{PS1}\end{equation}

\subsection{Scaling of the order parameter close to the transition \label{sub:Scaling-of-the}}

The expansion of the mean field equation (\ref{eq:BCS_equation})
would give an order parameter which scales as $B\sim\sqrt{g-g_{c}}$
close to the transition. This is the usual mean-field scaling. It
is modified by the strong fluctuations of local ordering fields present
in our problem. As we discuss in section \ref{sub:Replica-symmetric-regime}
this modification is present even in the some range of parameters
in replica symmetric phase not too far from RSB transition and, of
course, in the whole low temperature regime where RSB happens. We
shall analyze these two cases successively.

\subsubsection{Replica symmetric regime at large $K$.\label{sub:Replica-symmetric-regime}}

The nonlinear recursion relation (\ref{eq:mapping_Kfinite}) leads
to the equation for the average value \begin{equation}
\langle B\rangle=g\int dBP(B)\int_{0}^{1}d\xi\frac{B}{\sqrt{B^{2}+\xi^{2}}}\tanh(\beta\sqrt{B^{2}+\xi^{2}})\label{B01}\end{equation}
 At $T>T_{RSB}$, or $K>K^{RSB}(g)$ defined in Eq.(\ref{eq:K_RSB}),
the transition temperature is determined by Eq.(\ref{gc_naive}).
We assume now that, just below the transition line $T_{c}(K,g)$,
the distribution $P(B)$ has a scaling form $P(B)=B_{0}^{-1}\Phi(B/B_{0})$,
with $B_{0}$ vanishing at $T\to T_{c}^{-}$.

We begin by repeating the usual mean field arguments. Assuming that
$\Phi(x)$ decays faster than $1/x^{4}$, so that the integral $\int_{0}^{\infty}dx\Phi(x)x^{3}<\infty$,
we can expand \[
\frac{\tanh(\beta\sqrt{B^{2}+\xi^{2}})}{\sqrt{B^{2}+\xi^{2}}}\simeq\frac{\tanh(\beta\xi)}{\xi}-\frac{B^{2}}{2\xi^{4}}\left(\xi\tanh(\beta\xi)-\frac{\beta\xi^{2}}{\cosh^{2}(\beta\xi)}\right)\]
 We use the equation (\ref{gc_naive}) for the critical point to reduce
(\ref{B01}) to the form \begin{equation}
B_{0}\int dxx\Phi(x)\left(\frac{g}{g_{c}}-1\right)=B_{0}^{3}\int dx\Phi(x)x^{3}\int_{0}^{1}d\xi\frac{1}{2\xi^{4}}\left(\xi\tanh(\beta\xi)-\frac{\beta\xi^{2}}{\cosh^{2}(\beta\xi)}\right)\label{B02}\end{equation}
 which leads to the usual mean-field scaling $B_{0}\sim\sqrt{g-g_{c}}$.

Clearly, the crucial assumption used in this solution is that the
third moment of $P(B)$ (i.e. $\int\Phi(x)x^{3}dx$) is finite. It
breaks down if $P(B)$ decays at large $B$ like $B^{-1-a}$ with
$1<a<3$. Notice that in this regime the mean value of $B$ is finite,
therefore it belongs to the replica symmetric phase from the point
of view of the equations giving the critical temperature discussed
in section \ref{sub:Phase-diagram}. The behavior of $P(B)$ at large
$B$ can be studied by the Laplace transform method developed in section
\ref{sub:Distribution-function-of}. More precisely, we assume that
the Laplace transform at small $s$ has a form $1-\bar{B}s-As^{a}$
with $1<a<3$, insert it in the general recursion equation (\ref{eq:Laplace_recursion})
and solve the resulting equations for the coefficient $A$ and the
exponent $a$. The computation shows that the exponent $a$ is the
solution $a>1$ of the equation $f(a)=f(1)$. The mean field scaling
holds if and only if this solution satisfies $a>3$. This happens
when \begin{equation}
K>K_{3}=g_{c}^{3/2}\left[\int_{0}^{1}d\xi\left(\frac{\tanh(\beta\xi)}{\xi}\right)^{3}\right]^{1/2}\end{equation}
 The result $1-\bar{B}s-As^{a}$ for Laplace transform implies that
in the regime $K^{RSB}<K<K_{3}$ the distribution $P(B)$ on the critical
line $T=T_{c}(g)$ decays as $CB_{0}^{a}/B^{1+a}$ with $a<3$ and
$C\sim1$. In this case the equation (\ref{B02}) for the order parameter
is replaced by \begin{equation}
B_{0}(g-g_{c})=CB_{0}^{a}\int_{0}^{\infty}dxx^{-a}\int_{0}^{1}d\xi\left(\frac{\tanh(\beta\sqrt{B_{0}^{2}x^{2}+\xi^{2}})}{\sqrt{B_{0}^{2}x^{2}+\xi^{2}}}-\frac{\tanh(\beta\xi)}{\xi}\right)\label{B03}\end{equation}
 This implies the anomalous scaling of the order parameter near the
critical line: \begin{equation}
B_{0}(g)\propto(g-g_{c})^{1/(a-1)}\label{eq:B0(g)}\end{equation}
 Note that when $K\rightarrow K_{3}$ the exponent $a\to3$ so that
$B_{0}(g)$ dependence reduces to the usual square-root singularity.
At the opposite end of the interval $K^{RSB}<K<K_{3}$, the exponent
$\frac{1}{a-1}$ in Eq.(\ref{eq:B0(g)}) diverges because $a\to1$
as $K\rightarrow K^{RSB}$ .

\subsubsection{Regime of broken replica symmetry.\label{sub:Regime-of-broken}}

The RSB transition affects dramatically the scaling of the field in
the ordered phase in the vicinity of the transition. As we have seen
in subsection \ref{sub:Replica-symmetric-regime}, in order to derive
this scaling we need to know the distribution of fields $P(B)$ induced
by the mapping (\ref{eq:mapping_Kfinite}). As discussed in section
\ref{sub:Distribution-function-of} the expansion of its Laplace transform
shows that exactly on the transition line $g=g_{c}(K,T)$, in the
RSB regime, the distribution function $P(B)$ decays at large $B$
as $P(B)\propto1/B^{1+m}$. Here $m<1$ is the RSB parameter identified
in our analysis of the directed polymer problem as the value of $x$
where $f(x)$ is minimal.

For $g$ larger than $g_{c}$ but close to the transition, a small
typical field, $B_{0}$, appears. At $B\gg B_{0}$ the usual scaling
arguments show that the distribution $P(B)$ retains the same scaling
form (\ref{eq:P(B)_scaling_form}) that it acquired at $g_{c}$. As
we show below this scaling law is cut off from above by $B_{max}\simeq g/K\gg B_{0}$\,
, beyond which $P(B)$ decays rapidly. The power law behavior of $P(B)$
in a wide range of fields allows one to find analytically the typical
field $B_{0}$ from the following reasoning. Because the integral
$\int P(B)B^{m}dB\sim\ln(B_{max}/B_{0})$ is logarithmic, the main
contribution to the expectation value of $B^{m}$ comes equally from
a very broad range of fields $B_{max}\gg B\gg B_{0}$. Because of
this wide range of fields contributing to the expectation value, the
$m$'th power of (\ref{eq:mapping_Kfinite}) is dominated by the largest
term in the sum, so that \begin{equation}
B_{i}^{m}\simeq\sum_{k=1}^{K}\left(\frac{g}{K}\frac{B_{k}}{\sqrt{B_{k}^{2}+\xi_{k}^{2}}}\tanh\beta\sqrt{B_{k}^{2}+\xi_{k}^{2}}\right)^{m}\ .\label{eq:B^m_mapping}\end{equation}
 Similar arguments of dominance of the largest term justify the cutoff
at $B\simeq g/K\gg B_{0}$. Because each term in the sum (\ref{eq:mapping_Kfinite})
is cut off by $B_{max}=g/K$ and the sum is dominated by the largest
term, the power-law behavior (\ref{eq:P(B)_scaling_form}) is effectively
cut off by $B_{max}$, while at $B>B_{max}$ it decreases exponentially.

Averaging the approximate mapping for $B^{m}$ (\ref{eq:B^m_mapping})
we get the equation: \begin{equation}
\langle B^{m}\rangle=K\left(\frac{g}{K}\right)^{m\ }\int_{0}^{\infty}P(B)B^{m}dB\ \int_{0}^{1}d\xi\left(\frac{\tanh(\beta\sqrt{B^{2}+\xi^{2}})}{\sqrt{B^{2}+\xi^{2}}}\right)^{m}\label{B1}\end{equation}
 To proceed further we write the equation for the transition line,
$g_{c}=Ke^{-f(m)}$, in the form similar to Eq.(\ref{B1}): \begin{equation}
\left(\frac{g}{g_{c}}\right)^{m}\langle B^{m}\rangle=K\left(\frac{g}{K}\right)^{m}\int_{0}^{1}d\xi\left(\frac{\tanh(\beta\xi)}{\xi}\right)^{m}\int_{0}^{\infty}P(B)B^{m}dB\label{B2}\end{equation}
 and subtract (\ref{B2}) from (\ref{B1}):

\begin{multline}
[1-\left(\frac{g}{g_{c}}\right)^{m}]\langle B^{m}\rangle=K\left(\frac{g}{K}\right)^{m}\int_{0}^{\infty}P(B)B^{m}dB\\
\times\int_{0}^{1}d\xi\left\{ \left(\frac{\tanh(\beta\sqrt{B^{2}+\xi^{2}})}{\sqrt{B^{2}+\xi^{2}}}\right)^{m}-\left(\frac{\tanh(\beta\xi)}{\xi}\right)^{m}\right\} \label{eq:B^m_equation}\end{multline}
 As will be clear below, the qualitative properties of the solution
are the same at zero and finite temperatures. So, to simplify the
formulas we focus on the vicinity of the zero-temperature quantum
critical point $T_{c}=0$, $g=g_{c}(K)$. In this regime we can replace
all $\tanh(\beta u)$ functions by 1 and get \begin{equation}
[1-\left(\frac{g}{g_{c}}\right)^{m}]\langle B^{m}\rangle=K\left(\frac{g}{K}\right)^{m}\gamma(m)\langle B\rangle\label{B2a}\end{equation}
 where \begin{equation}
\gamma(x)=\int_{0}^{\infty}dt\left[\frac{1}{t^{x}}-\frac{1}{(t^{2}+1)^{x/2}}\right]\ .\label{B3a}\end{equation}
 In the large $K$ regime which is of main interest to us, $x\rightarrow1$,
and $\gamma(x)\to(1-x)^{-1}$. Using our \emph{Ansatz} for the distribution
function, valid in the broad range of $B_{max}\gg B\gg B_{0}$, we
find: \begin{equation}
\left[1-\left(\frac{g}{g_{c}}\right)^{m}\right]B_{0}^{m}\ln\frac{B_{max}}{B_{0}}=K\left(\frac{g}{K}\right)^{m}\frac{\gamma(m)}{1-m}\ CB_{0}^{m}B_{max}^{1-m}\label{B3}\end{equation}
 Here, and in the following, $C$ denotes a numerical constant of
order 1. Using the estimate for $B_{max}$ discussed above we get
\begin{equation}
B_{0}(g)\simeq e^{-1/(eg_{c})}\exp\left[-\frac{C\gamma(m)}{(g/g_{c})^{m}-1}\right]\label{B4}\end{equation}
 For $\epsilon=g/g_{c}-1\ll1$ we expand the exponent in Eq.(\ref{B4})
and find \begin{equation}
B_{0}(g)\simeq e^{-1/(eg_{c})}\exp\left[-\frac{C\gamma(m)}{m(g-g_{c})}\right]\label{B5}\end{equation}

This gives the typical value $B_{0}$ of the order parameter close
to the transition; in contrast to quantum critical points in clean
systems the behavior $B_{0}(g)$ displays an essential singularity.
This result becomes even more surprising when one compares $B_{0}$
to the critical temperature $T_{c}(g)=C(g-g_{c})^{1/(eg)}$ (which
follows from Eq.(\ref{eq:Tc(K)})) because it implies that the \textit{typical}
order parameter at $T=0$ is \textit{much smaller} than the transition
temperature at the same value of $g$.

Eq. (\ref{B4}) gives the dependence of the order parameter as a function
of the interaction constant at very low temperatures for $g$ close
to $g_{c}$. Similar arguments give the dependence of the order parameter
at $g=g_{RSB}$ as function of temperature in the vicinity of transition:

\begin{equation}
B_{0}(T)\simeq e^{-1/(2g_{RSB})}\exp\left[-\frac{C}{2g_{RSB}\ln(T_{c}/T)}\right]\end{equation}

\begin{figure}[ht]
 \includegraphics[width=7cm]{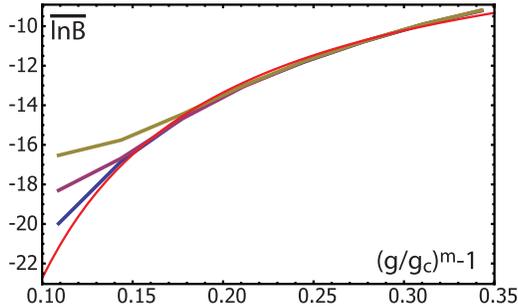} \caption{The typical order parameter vanishes with an essential singularity
at the superconductor-insulator transition. Upper curves show the
logarithm of the typical fields (defined by $\ln B_{typ}=\overline{\ln B}])$
obtained by population dynamics with small external fields: $h=10^{-8},10^{-9},10^{-10}$
(from top to bottom). The lower curve shows the fit to $1/((g/g_{c})^{m}-1)$
dependence expected from (\protect\ref{B4}). }

\label{fig:LogHtyp_K=00003D4} 
\end{figure}

Because this analytical derivation relies on our Ansatz for the distribution
function, we have checked the scaling behavior (\ref{B4}) numerically.
We solve the self consistent equation (\ref{eq:PdeB_Kfinite}) for
$P(B)$ using the population dynamics algorithm developed in work
\cite{MezardParisi2001}. In a nutshell, this method amounts to approximating
$P(B)$ by a population of fields, which is sampled by a Monte Carlo
method iterating Eq.(\ref{eq:mapping_Kfinite}). In order to tame
the effects of rare fluctuations, it is numerically convenient to
put the system in a small external field, using the mapping (\ref{Bext})
defined below in the Section~\ref{sub:Susceptibility-in-the}. The
results are reliable (as can be checked from the fact that they do
not depend on the external field at small enough field) when $g$
is not too close to $g_{c}$. The results for $K=4$ (for which the
zero-temperature critical coupling is $g_{c}\simeq0.1$ and $m\simeq0.73$)
are shown in Fig. \ref{fig:LogHtyp_K=00003D4} where we plotted $\overline{\log B}$
as a function of $(g/g_{c})^{m}-1$ for $K=4$ and its fit to a $[(g/g_{c})^{m}-1]^{-1}$
dependence: $\overline{\log B}=A+B[(g/g_{c})^{m}-1]^{-1}$ with $A=-4.0$,
$B=1.9$. This value of $B$ is close to the one expected from (\ref{B4})
because for this value of $K$ one has $\gamma(m)\approx2.8$.

\begin{figure}[ht]
 \includegraphics[width=7cm]{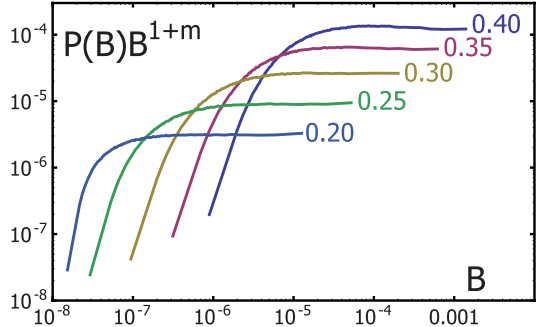} \caption{Distribution function, $B^{1+m}P(B)$ for $K=4$ and different values
of $(g-g_{c})/g_{c}$ shown next to each curve. As expected, $B^{1+m}P(B)$
is field-independent in a range of fields which is wide in logarithmic
scale, and becomes wider as the transition is approached.}

\label{fig:P(h)_K=00003D4} 
\end{figure}

The computations above are based on the assumption that the distribution
function retains its power-law dependence in the ordered phase in
a wide range of fields. We have checked this assumption directly using
population dynamics. The result is shown in Fig. \ref{fig:P(h)_K=00003D4}.
One observes that $P(B)B^{1+m}$ develops a plateau that becomes wider
on a logarithmic scale when one approaches the quantum critical point.

\subsection{Leading correction in $1/Z$ to the RSB solution. \label{sub:Leading-correction-in}}

As we explained in section \ref{sub:Cavity-mapping}, the full cavity
Hamiltonian (\ref{eq:H^cav}) can be approximately replaced by the
simpler Hamiltonian (\ref{eq:H^cav-MF}) in which the dynamics variables
$\sigma_{j}^{x}$ were replaced by their averages $\left\langle \sigma_{j}^{x}\right\rangle $.
This approximation ignores dynamic quantum correlations between the
spin at site $i$ and its neighbors at site $j$ . These correlations
might become important when the energies of these two sites are close,
so that $\left|\xi_{i}-\xi_{j}\right|\sim g/K$ . Such resonance conditions
occur with a probability $p\sim g/K$ which is small at large $K$.
To check that these and other corrections are not relevant at modest
values of $K$ we have performed a population dynamics simulation
of the full cavity mapping (\ref{eq:H^cav}). At each step of this
simulation we start with the population of $N$ spins characterized
by energies $\xi_{k}$ and fields $B_{k}.$ We randomly choose $N$
sets of $K$ spins from this population, add to each of them an additional
spin with random energy $\xi$ and diagonalize the corresponding Hamiltonian
(\ref{eq:H^cav}) to determine the effective value of the field acting
on the additional spin. This gives a new set of spins with energies
$\xi_{k}$ and fields $B_{k}.$ A typical simulation involved $10^{4}$
spins and $10^{4}$ steps which is sufficient to get the convergence
of the typical field at $(g-g_{c})/g_{c}>0.2$.

\begin{figure}
\includegraphics[width=4in]{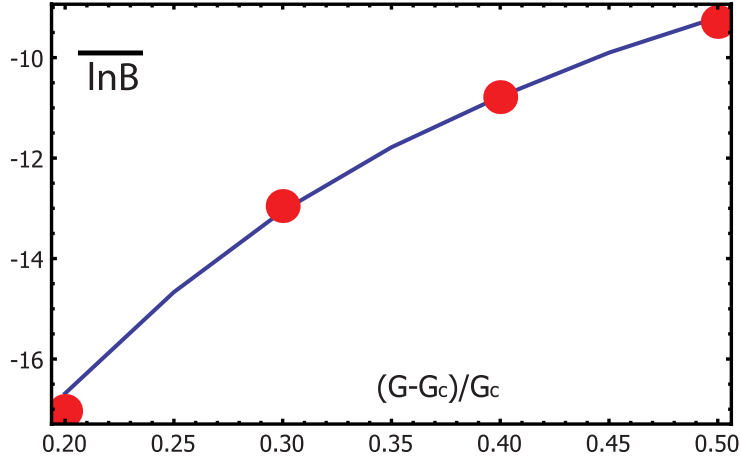} \caption{Comparison between the typical fields (measured by the harmonic mean)
computed with the population dynamics of the full cavity Hamiltonian
(\ref{eq:H^cav}) and the simplified one (\ref{eq:H^cav-MF}). The
line corresponds to the results of the simplified mapping, the dots
to the mapping given by the full cavity Hamiltonian. }

\label{Fig:FullandMF} 
\end{figure}

In Fig. \ref{Fig:FullandMF} we compare the results of the full diagonalization
with the simplified mapping for $K=4.$ As one can see from these
results, even at these modest value of $K$ the results of the full
cavity mapping are indistinguishable from the simplified mapping.
Qualitatively it means that the important physics driving this transition
is the appearance of wide distribution of fields, and this effect
is not sensitive to quantum correlations neglected by the simplified
mapping.

\subsection{Susceptibility in the disordered phase. \label{sub:Susceptibility-in-the}}

We now study the approach to the quantum critical point coming from
the insulating phase. In usual phase transitions the order parameter
induced by a small external field diverges at the transition, corresponding
to a divergent linear susceptibility. We shall see that the situation
is very different here: there is no diverging susceptibility.

We focus on the insulating phase at $T=0$ and $g<g_{c}$. We apply
a very small uniform external field $h$ to all sites. Repeating the
same arguments as before we arrive at the cavity-mean-field mapping
\begin{equation}
B_{i}=\frac{g}{K}\sum_{k=1}^{K}\frac{B_{k}}{\sqrt{B_{k}^{2}+\xi_{k}^{2}}}+h\ .\label{Bext}\end{equation}
 For very small $h$, in linear response, the induced fields are also
small. We can thus neglect the non-linearity in this equation at $g<g_{c}$.
In the resulting linear equation we can rescale the variables $B_{i}=b_{i}h$
and get \begin{equation}
b_{i}=\frac{g}{K}\sum_{k=1}^{K}\frac{b_{k}}{|\xi_{k}|}+1\label{b_i}\end{equation}
 All properties of the solution are contained in the distribution
function $P(b)$ generated by the mapping (\ref{b_i}). In order to
study this function, it is convenient to introduce the Laplace transform
$\mathcal{P}(s)=\int P(b)\exp(-sb)db$ . It satisfies a self-consistent
equation that can be derived directly from the mapping (\ref{b_i}):
\begin{equation}
\mathcal{P}(s)=e^{-s}\left[\int_{0}^{1}d\xi\mathcal{P}\left(\frac{g}{K}\frac{s}{\xi}\right)\right]^{K}\label{P(s)_equation}\end{equation}
 We assume the existence of $\mathcal{P}(s)$ and derive some of its
properties. The behavior at small $s$ is easily found: if we look
for a solution in the form $\mathcal{P}(s)=1-as^{\mu}$, we get the
equation for the exponent $\mu$:\[
K^{1-\mu}g^{\mu}=1-\mu\]
 This equation is equivalent to the equation for the RSB parameter
$m$ at $g=g_{c}$, see Eq.(\ref{eq:Kc}). Therefore $\mu=m$.

\begin{figure}[ht]
 \includegraphics[width=7cm]{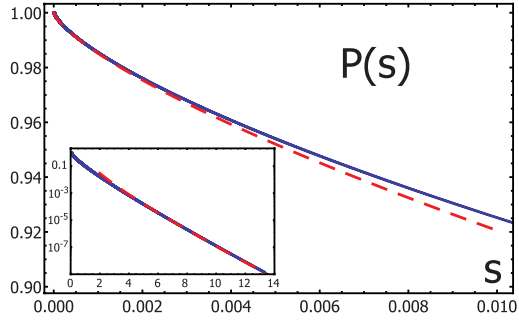} \caption{Laplace transform of the distribution function produced by a small
external field at $g=g_{c}$ for $K=4$. Main panel shows the regime
of small $s$, the inset shows the behavior at large $s$. The dashed
lines show the asymptotic behavior at small and large $s$ discussed
in the text.}

\label{fig:P(s)_K=00003D00003D00003D4} 
\end{figure}

When $s\to\infty$ we look for a solution in the form $\mathcal{P}(s)=cs^{b}\exp(-as)$.
Inserting it into the equation (\ref{P(s)_equation}) we get, for
$s\gg K/g$: \begin{equation}
a=\frac{1}{1-g}\qquad b=\frac{K}{K-1}\qquad c=\left(\frac{1}{(1-g)}\ \left(\frac{K}{g}\right)^{1/(K-1)}\right)^{\frac{K}{K-1}}\label{Large s asymptotics}\end{equation}
 In order to prove the existence and stability of this solution we
solved the equation (\ref{P(s)_equation}) numerically. In order to
check the analytical asymptotic we need to have the solution in a
broad range of parameter $s$. For this reason we have transformed
the equation (\ref{P(s)_equation}) by introducing the variables $s=e^{\zeta}$,
$\xi=e^{-\tau}$:

\begin{equation}
\widetilde{\mathcal{P}}(\zeta)=e^{-\exp\zeta}\left[\int_{0}^{\infty}d\tau\exp(-\tau)\widetilde{\mathcal{P}}\left(\log\left(\frac{g}{K}\right)+\zeta+\tau\right)\right]^{K}\end{equation}
 and solved the resulting equation by iteration for $K=4$ and $g=g_{c}=0.0996$.
The result is displayed in Fig. \ref{fig:P(s)_K=00003D00003D00003D4}.
The low $s$ asymptotic is indeed $\mathcal{P}(s)=1-as^{m\text{ }}$,
with the correct value of $m=1-0.1\cdot e$. The asymptotic behavior
at large $s$ is more tricky because the exponential dependence $\mathcal{P}(s)=cs^{b}\exp(-as)$
with parameters given in Eq.(\ref{Large s asymptotics}) is expected
to occur when $gs/K\gg1$; this condition implies that in this regime
$\mathcal{P}(s)\ll10^{-20}$ which is difficult to access numerically.
In the transient regime of $K/g\gg s\gg1$ the numerical solution
fits exponential dependence with somewhat different parameters, $a=1.2$
and $b=-1.7$.

\begin{figure}[ht]
 \includegraphics[width=7cm]{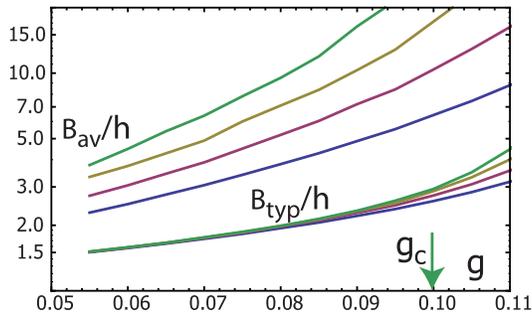} \caption{Average and typical fields induced by an external field at $g<g_{c}$
and $K=4$. The lower set of curves shows the typical response $B_{typ}/h=\exp(\overline{\log B})/h$
for external fields $h=10^{-6},10^{-7},10^{-8}$ and $10^{-9}$ (bottom
to up). The upper set of curves shows the average response $B_{av}/h=\overline{B}/h$
for the same values of the external field. The curves show no singularity
at the critical value of $g$ indicated by arrow. As discussed in
the text, the average response to the external field is controlled
by the far tail of the distribution function and is non-linear, so
the ratio $B_{av}/h$ grows at $h\rightarrow0$.}

\label{fig:Hav_and_Htyp_K=00003D00003D00003D4} 
\end{figure}

The existence of a stable solution $\mathcal{P}(s)$ of the equation
(\ref{P(s)_equation}) implies a number of unusual properties of this
$T=0$ quantum phase transition. A small external field $h$ induces
some non-zero fields $B_{i}$. Let us compute the average $\overline{B_{i}^{x}}$.
It can be expressed through the Laplace transform by \[
\overline{B^{x}}=\frac{x}{\Gamma(1-x)}\int_{0}^{\infty}ds\frac{1-\mathcal{P}(s)}{s^{1+x}}\,.\]
 The behavior $\mathcal{P}(s)=1-as^{m\text{ }}$ implies that for
$x<m$ the average $\overline{B^{x}}\sim h^{x}$. In particular, the
typical induced order parameter is of the order of the external applied
field: $\exp(\overline{\log B})\sim h$. But the moments $\overline{B^{x}}$
with $x>m$, and in particular the mean $\overline{B}$, are divergent
at the level of the linear response to $h$. The non-linearity of
the mapping neglected in (\ref{b_i}) cuts off this divergence at
$B\sim g/K$. One can therefore expect a non-linear response $\overline{B^{x}}\sim(g/K)^{x-m}h^{m}$
when $x>m$. These results show that the response to an external field,
computed at $g<g_{c}$, has no singularity at the transition. This
behavior is totally different from the one in usual phase transitions.
We illustrate this by Fig. \ref{fig:Hav_and_Htyp_K=00003D00003D00003D4}
which shows the average and typical fields induced by a small external
field at $g<g_{c}$. 

\subsection{Spatial scale of inhomogeneities of the order parameter.\label{sub:Spatial-scale-of}}

Close to the transition the spatial scales beyond which the system
is uniform become very large. In particular, at temperatures below
that of replica symmetry breaking, the susceptibility is dominated
by a single path, as discussed above, in Section~\ref{sub:The-directed-polymer},
implying that the system is essentially non-uniform at all length
scales. The goal of this section is to compute the characteristic
scales at which the system becomes uniform in the ordered state.

The non-uniformity at short scales is related to the fact that close
to the transition the order parameter in the infinite system is power-law
distributed in an exponentially wide range, from $B_{0}$ to $g/K$.
In contrast, at a given site the order parameter has some value which
changes by a factor $O(1)$ in the vicinity of this site. Thus, at
short scales the order parameter acquires values of the same order
of magnitude, whilst the full distribution function is formed only
at large scales.

In order to describe this physics quantitatively, we write down the
equations for the spatial evolution of the Laplace transform of the
distribution function upon iteration on the Bethe lattice: \begin{equation}
\mathcal{P}_{n+1}(s)=\left[\int_{0}^{1}d\xi\int dbP_{n}(B)\exp\left(-\frac{g}{K}\frac{sB}{\sqrt{\xi^{2}+B^{2}}}\right)\right]^{K}\label{eq:P_{n+1}(s)_general}\end{equation}
 The crucial feature of the stationary solution of this equation is
a power-law dependence of the distribution function in the exponentially
wide range of $s$: $1\ll s\ll1/B_{0}$. In this range we can neglect
the non-linear ($B^{2}$) term in the square root of (\ref{eq:P_{n+1}(s)_general}),
the same is true for a general (non-stationary) solution of equation
(\ref{eq:P_{n+1}(s)_general}) in this parameter range. This allows
to reduce the equation to the evolution of Laplace transforms: \begin{equation}
\mathcal{P}_{n+1}(s)=\left[\int_{0}^{1}d\xi\,\mathcal{P}_{n}\left(\frac{g}{K}\frac{s}{\mbox{\ensuremath{\xi}}}\right)\right]^{K}\label{eq:P_{n+1}(s)}\end{equation}

The stationary solution of this equation was discussed above, in Section~\ref{sub:Distribution-function-of}.
Here we need to find the spatial scales (i.e. the number of iterations)
at which this stationary solution emerges for $P_{0}(B)$ that corresponds
to a particular value of the field, $B=B_{1}$, i.e. $\mathcal{P}_{1}(s)=\exp(-sB_{1})$.

The initial stages of evolution lead to the distribution function
that has many features of the stationary solution given by Eqs.(\ref{eq:P(B)_scaling_form},\ref{PS1}),
$\mathcal{P}(s)=1-\alpha(sB_{0})^{m}$, in particular it becomes close
to unity in a broad range of $s.$ The final spreading over the whole
range $1\ll s\ll1/B_{0}$ and thus the spatial scale at which the
stationary solution is realized can be described by the linearized
equation for $\phi(s)=1-\mathcal{P}(s)$: \[
\phi_{n+1}(s)=K\int_{0}^{1}d\xi\phi_{n}\left(\frac{g}{K}\frac{s}{\mbox{\ensuremath{\xi}}}\right)\]
 The evolution described by this equation approaches slowly the stable
stationary solution $\phi_{\infty}(s)=\alpha(sB_{0})^{m}$ found before.
As we shall see below, this evolution is similar to a diffusion equation,
so the total number of steps ({}``time'') needed for this evolution
is controlled by the final spreading of the distribution function
(the smaller $B_{0}$, the more iterations it takes to reach the stationay
solution). We can study this convergence towards $\phi_{\infty}$
by assuming smooth deviations: $\phi_{n}(s)=y(\ln s)\phi_{\infty}(s)$
, where $y(\zeta)$ is a slow function of its variable. We get \begin{equation}
y_{n+1}(\zeta)=K\int_{0}^{1}d\xi\,\left(\frac{g}{K\xi}\right)^{m}\, y_{n}\left(\ln\frac{g}{K\xi}+\zeta\right)\label{yz}\end{equation}
 The integral over $\xi$ in Eq.(\ref{yz}) is dominated by $\xi\sim1$,
whereas $|\ln(g/K)|\approx1/(eg_{c})$. The assumption that $y(\zeta)$
is a slow function on the scale of $(eg_{c})^{-1}$ allows us to expand
$y_{n}\left(\ln\frac{g}{K\xi}+\zeta\right)$ in powers of $\ln\frac{g}{K\xi}$
. Carrying this expansion up to the second order we get \begin{equation}
y_{n+1}(\zeta)=uy_{n}(\zeta)+v\frac{dy_{n}}{d\zeta}+D\frac{d^{2}y_{n}}{d\zeta^{2}}\label{eq:y_n+1(zeta)}\end{equation}
 The coefficient $u=K\int_{0}^{1}d\xi(g/K\xi)^{m}$ is equal to 1
when $g=g_{c}$, and $\nu=K\int_{0}^{1}d\xi(g/K\xi)^{m}\log(g/K\xi)=0$
because of the stationarity condition which gives $m$. Notice that,
if this stationarity condition were not satisfied, one would get $\nu\neq0$,
implying a non-zero drift term in the equation (\ref{eq:y_n+1(zeta)}).
In this situation s stationary solution for the probability distribution
would be impossible. This argument proves that the existence of a
stationary solution of the recursion equation for the probability
distribution of the order parameter implies that the value of $m$
is fixed by the stationarity condition, as we found in the replica
analysis.

Altogether the integral equation (\ref{yz}) reduces to the equation
of diffusive evolution: \begin{equation}
y_{n+1}(\zeta)-y_{n}(\zeta)=D\frac{d^{2}y_{n}(\zeta)}{d\zeta^{2}}\label{yz1}\end{equation}
 with a diffusion coefficient \begin{equation}
D=\frac{1}{2}\ln^{2}\frac{g}{K}\label{eq:D}\end{equation}
 The longest relaxation {}``time'' of this diffusive motion on the
interval $(0,\ln\frac{1}{B_{0}})$ is given by \begin{equation}
N=\frac{4\ln^{2}B_{0}}{\pi^{2}D}\label{eq:N}\end{equation}
 This relaxation {}``time'' is actually equal to the correlation
length of our problem. Close to the transition, $B_{0}$ goes to zero
as in Eq.(\ref{B5}), and therefore the length scale at which the
system becomes essentially uniform diverges as $N\propto(g-g_{c})^{-2}$.

\section{Width of the levels in the insulating state.\label{sec:Width-of-the-levels}}

In the disordered phase the average value of the transverse field
is zero. However the fluctuations of this field may be important.
The main physical effect of these fluctuations is the broadening of
the local levels that corresponded to $\sigma_{i}^{z}=\pm1$ in the
$g\to0$ limit. We shall study this broadening both at $T=0$ and
at $T\neq0$. Note that level broadening at $T=0$ is a rather complex
phenomenon. It implies that a local excitation of spin $i$ at frequency
$\omega=2\xi_{i}$ decays. By energy conservation such a decay implies
the excitation of some other spins. This cannot happen in a finite
system because the energy of the spins are discrete and random. So
the level broadening effect can only appear in an infinite system,
where the excitation can propagate to an infinite number of other
spins. We will show here that the broadening of levels in the insulating
phase appears as a phase transition.

\subsection{Propagation of time-dependent perturbations, mobility edge.\label{sub:Propagation-of-time-dependent}}

In order to study infinite systems consistently, we adopt an approach
similar to the one developed above for the study of the transition
into the ordered state. Namely, we consider an infinite Bethe lattice
which is very weakly coupled to the environment at its boundary, study
the effective level width at a distance $L$ from the boundary and
take the limit $L\rightarrow\infty$. Thus we add to the Hamiltonian
(\ref{H_B}) the boundary term $H_{env}=\sum_{i}(\sigma_{j}^{+}x_{j}(t)+h.c.)$
where $x_{j}(t)$ are \emph{dynamical} fields, generated by the environment,
characterized by a spectral function $S(\omega)$. In the leading
order of the perturbation theory in $g/K$, the effect of these fields
on the spin $0$ at a distance $L$ from the boundary follows from
the Fermi Golden rule. Imagine that the environment of a spin $j$
on the boundary induces a perturbation of frequency $\omega$ in the
form $\sigma_{j}^{+}x_{j}(\omega)e^{-i\omega t}+h.c.$. This perturbation
induces a matrix element between the two states of the spin 0 corresponding
to $\sigma_{0}^{z}=\pm1$. This matrix element appears only in the
$L^{th}$ order of the perturbation theory in $g$ and is equal to
$x_{j}(\omega)\prod_{k}[(2g/K)/(\omega-2|\xi_{k}|)]$, where the index
$k$ runs along the path connecting spin $0$ and the spin $j$ at
the boundary. Thus, in this approximation, the application of the
golden rule to the relaxation rate of the spin 0 gives, at zero temperature:
\begin{equation}
\Gamma_{0}=\sum_{P}\prod_{k\in P}\left[\frac{2g/K}{\omega-2|\xi_{k}|}\right]^{2}S(\omega)\label{eq:Gamma_Linear_Mapping}\end{equation}
 where the sum runs over all paths connecting spin $0$ to the spins
at the boundary at distance $L$. This equation is valid provided
that all fractions inside the product remain small, the usual condition
for the validity of perturbation theory. It should be modified when
some of the fractions get large, but as we will see below these cases
are so rare that these modifications are irrelevant.

The study of the spontaneous emergence of a finite width can be done
using the same directed polymer technique that we used in the previous
section. Before we get into the details of this derivation it is useful
to summarize the results that we shall obtain. When $g<g_{d}(\omega,K)$
the levels have zero width, they are discrete. The curve $g=g_{d}(\omega,K)$
defines the spontaneous appearance of a finite width. $g_{d}(0,K)$
is equal to the (zero-temperature) critical coupling where the system
becomes superconductor: $g_{d}(0,K)=g_{c}(K)$. For finite $\omega$,
$g_{d}(\omega,K)<g_{d}(0,K)$; the minimal value of $g_{d}(\omega,K)$
as function of $\omega$ occurs in the middle of the band, at $\omega=1$.
At $g<g^{\ast}(K)=g_{d}(\omega=1,K)$ the relaxation rate is zero
for all states. This regime corresponds to the superinsulator introduced
in \cite{Basko2006}. For a given intermediate value of the coupling
$g^{\ast}(K)<g<g_{c}(K)$, the states in the middle of the band have
a finite width. They are separated from the states of zero width by
a critical energy $\omega_{d}(g,K)$ (obtained by inverting the function
$g_{d}(\omega,K)$) similar to the mobility edge of the non-interacting
problem.

We now derive these results using the mapping to directed-polymers.
The partition function of the directed polymer on the tree is now:
\begin{equation}
\Xi=\sum_{P}\prod_{k\in P}\left[\frac{2g/K}{\omega-2|\xi_{k}|}\right]^{2}\label{eq:Xi}\end{equation}
 The computation of $\overline{\log\Xi}$ is deduced from the evaluation
of the function \begin{equation}
f_{\omega}(x)=\frac{1}{x}\ln\left[K\int_{0}^{1}d\xi\left(\frac{2}{\left|\omega-2\xi\right|}\right)^{2x}\right]\ .\label{eq:f_omega(x)}\end{equation}
 In the directed polymer terminology this problem turns out to be
always in the low temperature phase corresponding to replica symmetry
breaking, where the main contribution comes from a very small number
of paths. This is due to the fact that $f_{\omega}(x)$ always has
a minimum at $x=b<1$. As a consequence, one finds, using the same
approach as in Sect.\ref{sub:The-directed-polymer}: $(1/L)\overline{\log\Xi}=f_{\omega}(b)+2\ln(g/K)$.

When $\omega=0$ one has $f_{\omega=0}(x)=2f(2x)$, where $f$ is
given in (\ref{eq:f(x)}). Therefore, in the whole insulating regime
$g<g_{c}$ one gets $(1/L)\overline{\ln\Xi}<0$. Thus, the relaxation
rate at very low frequencies decreases away from the boundary, in
the bulk of the sample the width of levels is zero.

Let us now study the case of non-zero frequencies. The critical line
at which a finite level width appears in the bulk of the sample is
given by the two equations $\frac{df_{\omega}}{dx}\left|_{x=b}\right.=0$
and $f_{\omega}(b)+2\ln(g/K)=0$ which read: \begin{eqnarray}
K\left(\frac{g}{K}\right)^{2b}\int_{0}^{1}d\xi\frac{1}{|\xi-\omega/2|^{2b}} & = & 1\label{eq:eq_g(omega)1}\\
K\left(\frac{g}{K}\right)^{2b}\int_{0}^{1}\frac{d\xi}{|\xi-\omega/2|^{2b}}\ln\frac{1}{|\xi-\omega/2|} & = & \ln\frac{K}{g}\label{eq:eq_g(omega)2}\end{eqnarray}
 These equations can be written explicitly: \begin{eqnarray}
K^{z}g^{1-z}\left[(1-\omega/2)^{z}+(\omega/2)^{z}\right] & = & z\label{eq:eq_g(omega)3}\\
K^{z}g^{1-z}\left[(1-\omega/2)^{z}\ln(1-\omega/2)+(\omega/2)^{z}\ln(\omega/2)\right] & = & 1-z\ln\frac{K}{g}\label{eq:eq_g(omega)4}\end{eqnarray}
 where we have introduced the notation $z=1-2b$.

We begin by finding the region of the parameters $(g,K)$ in which
the system of equations (\ref{eq:eq_g(omega)1},\ref{eq:eq_g(omega)2})
has a solution. Clearly, the solution of these equations expressed
as a function of $K_{g}(\omega)$ is symmetric under $\frac{\omega}{2}\rightarrow1-\frac{\omega}{2}$
at fixed $g$. The energy $\omega=1$ corresponds to the minimal value
of $K_{g}(1)=K^{\ast}(g)$. For $K<K^{\ast}(g)$ the equations (\ref{eq:eq_g(omega)1},\ref{eq:eq_g(omega)2})
have no solution, so at these values of the cavity degree $K$ all
energy levels have zero width in the bulk of the sample. At $\omega=1$
the system of equations (\ref{eq:eq_g(omega)3},\ref{eq:eq_g(omega)4})
can be solved explicitly: \begin{eqnarray}
K^{\ast}(g) & = & 2ge^{1/(2eg)}\label{eq:K*(g)}\\
z^{*}(g) & = & 2eg\label{eq:z*(g)}\end{eqnarray}
 Notice that for small $g$ the value $K^{\ast}(g)$ is exponentially
smaller than the value $K_{c}(g)$, given by (\ref{eq:K_c}), at which
superconductivity appears.

Next, we consider the region of small $\omega$. To solve the system
(\ref{eq:eq_g(omega)3},\ref{eq:eq_g(omega)4}) in this limit, we
first notice that at $\omega=0$ the solution of the equations gives
the critical point discussed in Sect.\ref{sub:Phase-diagram}: $K_{g}(0)=K_{c}(g)=ge^{1/(eg)}$
and $z(0)=eg$. At small $\omega\ll1$ and $g\ll1$ the exponent $z$
should remain small: $z\ll1$ which allows us to neglect terms linear
in $\omega$ in (\ref{eq:eq_g(omega)3},\ref{eq:eq_g(omega)4}): \begin{eqnarray}
e^{(z/eg)-1}\left[1+(\omega/2)^{z}\right] & = & \frac{z}{eg}\left(\frac{K_{c}}{K}\right)^{z}\label{eq:eq_small_omega1}\\
z\frac{(\omega/2)^{z}}{1+(\omega/2)^{z}}\ln(\omega/2) & = & 1-\frac{z}{eg}+z\ln\frac{K_{c}}{K}\label{eq:eq_small_omega2}\end{eqnarray}
 The second equation (\ref{eq:eq_small_omega2}) shows that $(1-z/eg)$
scales as $\omega^{z},$ so deviations of $z$ from its critical value
can be neglected in the first equation where terms linear in $(1-z/eg)$
cancel. This gives \begin{equation}
\omega_{d}(g,K)=2\left[\left(\frac{K_{c}(g)}{K}\right)^{eg}-1\right]^{1/(eg)}\label{eq:omega_d(g,K)}\end{equation}
 This reduces to a power law behavior of $\omega_{d}(g,K)$ when $K$
is close to $K_{c}$ and $\omega_{d}\ll1$: \begin{equation}
\omega_{d}(g,K)=2(eg)^{1/(eg)}\left(1-\frac{K}{K_{c}(g)}\right)^{1/(eg)}\label{eq:omega_d(g,K)_1}\end{equation}
 These results are illustrated by Fig. 8 which
shows $\omega_{d}(g=0.129,K)$ as function of $K$.

Note that the previous analysis shows that the region of small $\omega-2\xi$
gives a negligible contribution to $f_{\omega}$. This gives an \emph{a
posteriori} justification to the fact that we have neglected the non-perturbative
modifications of the equation (\ref{eq:Gamma_Linear_Mapping}) in
our study of the insulating phase.

\begin{figure}
\label{omega8}\includegraphics[width=4in]{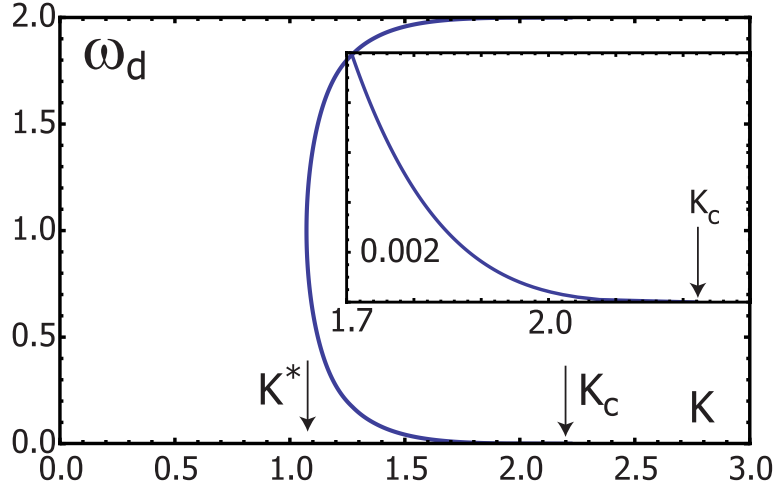}

\caption{Critical frequency, $\omega_{d}$, that separates zero-width levels
from the levels with non-zero width, plotted as a function of $K$
for $g=0.129$. This value of $g$ was chosen because it corresponds
to the experimentally relevant transition temperature at large $K$:
$T_{BCS}=0.001$ (see section \ref{sec:Consequences-for-experiments.}).
At $K<K^{*}$ all states have zero width. In the intermediate regime
$K^{*}<K<K_{c}$ states to the left of $\omega_{d}(K)$ line have
zero width (infinite decay time). The insert shows a blow-up region
around $K_{c}.$}

\end{figure}

\subsection{Scaling of the level width close to the transition. \label{sub:Scaling-of-the-level-width}}

The level widths $\Gamma(\omega)$ strongly depend on the frequency
for $\omega>\omega_{d}(g)$. The form of this dependence is important
for the study of low temperature properties discussed below. To find
$\Gamma(\omega)$ we need to rederive the equation (\ref{eq:Gamma_Linear_Mapping})
for the case of non-negligible $\Gamma(\omega)$. The computation
using the Keldysh formalism, which we shall explain in Section \ref{sub:The-effect-of-non-zero-temperature}
below, gives at zero temperature: \begin{equation}
\Gamma_{i}=(2g/K)^{2}\sum_{k(i)}\frac{\Gamma_{k}}{\left(\omega-2\xi_{k}\right)^{2}+\Gamma_{k}^{2}}\label{eq:Gamma_Nonlinear_mapping}\end{equation}
 As one expects, the decay rate of the state $k$ provides the cutoff
of the divergence at $\omega\rightarrow2\xi_{k}.$ This equation is
similar to the equation (\ref{eq:mapping_Kfinite}) for the fields
appearing in the superconducting phase. We are interested in the scaling
of the typical level width for $\omega>\omega_{d}(g,K)$ but close
to the transition.

We will analyze this scaling with the same method as in Sect.~\ref{sub:Scaling-of-the},
in the RSB region. The crucial ingredient is the distribution of widths
$W(\Gamma)$. We shall assume that it has a power law form with an
upper cutoff : \[
W(\Gamma)=\frac{\Gamma_{0}^{b}}{\Gamma^{b+1}}\mbox{{\ where\ }}\Gamma_{0}\leq\Gamma\leq\Gamma_{{\rm max}}\]
 The non-linear mapping (\ref{eq:Gamma_Nonlinear_mapping}) gives
a self-consistent equation for this distribution. Performing the same
steps as those used in the derivation of Eq.(\ref{B3}), we get: \begin{equation}
\left[\left(\frac{g}{g_{d}(\omega,K)}\right)^{2b}-1\right]\int d\Gamma\ W(\Gamma)\,\Gamma^{b}=\frac{K}{2}\left(\frac{2g}{K}\right)^{2b}\int d\Gamma\, W(\Gamma)\,\Gamma^{b}\,\gamma\left(2b,\frac{\omega}{\Gamma}\right)\label{Gamma21}\end{equation}
 where the function $\gamma(x,y)$, defined by: \begin{equation}
\gamma(x,y)=\int_{0}^{\infty}dt\left[\frac{1}{|t-y|^{x}}-\frac{1}{[(t-y)^{2}+1]^{x/2}}\right]\ ,\end{equation}
 reduces to $\gamma(x)$ when $y\to0$. The mapping (\ref{eq:Gamma_Nonlinear_mapping})
implies that values $\Gamma$ much larger than $g/K$ are very rare.
Indeed, such a value of $\Gamma$ at one step induces a much smaller
value at the next step. We thus assume $\Gamma_{{\rm max}}\sim g/K\gg\omega_{d}$.
Evaluating both sides of Eq. (\ref{Gamma21}) we find \begin{equation}
\left[\left(\frac{g}{g_{d}(\omega,K)}\right)^{2b}-1\right]\ln\frac{g}{K\Gamma_{0}}=\frac{K}{2}\left(\frac{2g}{K}\right)^{2b}\frac{\gamma(2b)}{1-2b}\Gamma_{max}^{1-2b}\equiv C\gamma(m)\label{Gamma22}\end{equation}

Close to the critical point $K=K_{d}(g)$, the exponent $2b\approx m=1-eg$.
Expanding $(g/g_{d}(\omega,K))^{m}-1$ at small $(\omega-\omega_{d})/\omega_{d}$
we find using the Eq.(\ref{eq:omega_d(g,K)_1}) that \begin{equation}
\left(\frac{g}{g_{d}(\omega,K)}\right)^{m}-1\approx(eg)^{2}\left(1-\frac{K}{K_{c}}\right)\frac{\omega-\omega_{d}(g,K)}{\omega_{d}(g,K)}\label{Gamma-g}\end{equation}
 Finally, we obtain for the typical level width: \begin{eqnarray}
\Gamma^{typ}(\omega)\simeq\Gamma_{0}(\omega) & \simeq & e^{-1/eg}\exp\left(-\frac{\omega_{1}(g,K)}{\omega-\omega_{d}(g,K)}\right)\label{eq:Gamma^typical(omega)}\\
\omega_{1}(g,K) & = & \frac{C\gamma(2b)}{(eg)^{2}}\frac{\omega_{d}(g,K)}{1-(K/K_{c})}\gg\omega_{d}(g,K)\label{eq:omega_1}\end{eqnarray}
 As we shall discuss below, this fast energy dependence of the level
width has important consequences for the low-temperature transport
properties. As before, the numerical coefficient $C\sim1$ cannot
be determined analytically because we do not know the precise value
of $\Gamma_{{\rm max}}$. We emphasize that, according to Eq.(\ref{eq:omega_1}),
the energy scale $\omega_{1}$ in the exponent is much larger than
the threshold energy $\omega_{d}$. This inequality $\omega_{1}\gg\omega_{d}$
is valid in the whole range of validity of Eq. (\ref{eq:omega_1}),
as long as $\omega_{d}\leq g/K$. Qualitatively, it leads to a very
sharp growth of the typical level width $\Gamma_{0}(\omega)$ right
above the threshold.

The estimate of the level width close to the quantum critical point,
$K_{c}(g)$, requires a special treatment because the expansion (\ref{Gamma-g})
and the final result (\ref{eq:Gamma^typical(omega)}) are valid only
for frequencies $\omega$ close to the threshold $\omega_{d}$ so
that $\omega-\omega_{d}\ll\omega_{d}$. Thus, they are not applicable
at the critical point $K=K_{c}(g)$ where both $\omega_{d}$ and $\omega_{1}$
vanish. An approximate expression for $\Gamma_{0}(\omega)$ in this
regime should be found differently. First, using the function $\omega_{d}(g,K)$
given by Eq. (\ref{eq:omega_d(g,K)_1}), we determine the inverse
function $g_{d}(\omega,K)$ for low $\omega$: \[
1-\frac{g_{d}(\omega,K)}{g_{c}}=\left(\frac{\omega}{2}\right)^{eg_{c}}\frac{1}{1-eg_{c}}\]
 Next, we substitute this expression into Eq. (\ref{Gamma22}) and
find \begin{equation}
\Gamma_{0}(\omega)=\frac{g_{c}}{K}\exp\left[-\Upsilon(g_{c})\left(\frac{2}{\omega}\right)^{eg_{c}}\right]\label{eq:Gamma_at_Kc}\end{equation}
 where $\Upsilon(g_{c})$ is some function of $g_{c}$ only. At $eg_{c}\ll1$
we find $\Upsilon(g_{c})=C(e^{2}g_{c})^{-1}$.

\subsection{The effect of a non-zero temperature on the level width. \label{sub:The-effect-of-non-zero-temperature}}

In this section we derive the cavity equations for the level width
within the Keldysh formalism. This allows to study the effect of a
non-zero but low temperature. A low temperature affects the relaxation
rate in several ways. First of all, it changes the occupation numbers
of the excited states and the ground states; this affects the perturbative
equation (\ref{eq:Gamma_Linear_Mapping}) and thus shifts the position
of the $\omega_{d}(g)$ line. This effect is however small at $T\ll1.$
A more important effect is that, in the intermediate phase $K^{*}(g)<K<K_{c}(g)$,
a non-zero temperature induces a small number of mobile excitations
with frequencies above $\omega_{d}(g,K)$. These excitations provide
a mechanism for a small but non-zero level broadening for the levels
even at very low frequencies, $\omega<\omega_{d}(g)$.

The effect of a small number of mobile excitations can be estimated
qualitatively by the following arguments. The excitation with energy
$E$ moves with a typical rate $\Gamma^{typ}(\omega)\sim\exp(-\omega_{1}/(E-\omega_{d}))$
from site to site. Thus, with exponential accuracy a typical site
sees a mobile excitation with energy $E$ passing by with rate $\exp(-\omega_{1}/(E-\omega_{d})-E/T).$
The dominant contribution to the relaxation of this site comes from
the energies $E=\omega_{d}(g)+\sqrt{\omega_{1}T}$. It results in
the temperature dependence $\Gamma\sim\exp(-2\sqrt{\omega_{1}/T}-\omega_{d}/T)$
that shows a crossover between a behavior  $\exp(-C/\sqrt{T})$
and an Arrhenius behavior $\exp(-C/T)$ as one goes away from the
critical point.

We shall now use the Keldysh formalism to support these qualitative
reasoning.

\subsubsection{Cavity equations for the relaxation rate: derivation with the Keldysh
formalism.}

As a first step, to be used later in the cavity approach, we begin
with a study of a reduced two-spin system where a fluctuating field
$\widehat{h}(t)$ is coupled to spin $\sigma_{1}$. The Hamiltonian
describing this system is \begin{equation}
H=-\xi_{0}\sigma_{0}^{z}-\xi_{1}\sigma_{1}^{z}-(2g/K)(\sigma_{0}^{+}\sigma_{1}^{-}+\sigma_{0}^{-}\sigma_{1}^{+})-\left[\widehat{h}(t)\sigma_{1}^{-}+\widehat{h}^{*}(t)\sigma_{1}^{+}\right]\label{eq:H_TwoSpin}\end{equation}
 We need to find the effective relaxation rate of spin $0$. For this
we employ the Jordan-Wigner transformation, using Fermion creation
operators $c_{0}^{\dagger}$ and $c_{1}^{\dagger}$: \begin{eqnarray}
\sigma_{0}^{+} & = & c_{0}^{\dagger}(1-2c_{1}^{\dagger}c_{1}),\;\sigma_{0}^{z}=2c_{0}^{\dagger}c_{0}-1\label{JordanWigner}\\
\sigma_{1}^{+} & = & c_{1}^{\dagger},\;\sigma_{1}^{z}=2c_{1}^{\dagger}c_{1}-1\end{eqnarray}
 The Keldysh action obtained after averaging over the environment
variables is: \begin{eqnarray}
S & = & \sum_{\alpha}\int(\overline{c}_{0\alpha}i\partial_{t}c_{0\alpha}+\overline{c}_{1\alpha}i\partial_{t}c_{1\alpha}-\alpha H[c])dt-\sum_{\alpha,\beta}\int C_{\alpha\beta}(t-t^{\prime})\overline{c}_{1\alpha}(t)c_{1\beta}(t^{\prime})dtdt^{\prime}\label{TwoSpinAction}\\
H[c_{\alpha}] & = & -(2\xi_{0}\overline{c}_{0\alpha}c_{0\alpha}+2\xi_{1}\overline{c}_{1\alpha}c_{1\alpha}+(2g/K)(\overline{c}_{0\alpha}c_{1\alpha}+\overline{c}_{1\alpha}c_{0\alpha}))\end{eqnarray}
 Here subscripts $\alpha,\beta=\pm$ are Keldysh indices corresponding
to fermions moving forth or back in time. It is convenient to rotate
all fermion vectors in the Keldysh space according to the Larkin-Ovchinnikov
prescription (for details of Keldysh formalism see the recent detailed
review \cite{KamenevReview}). For each $i\in\{0,1\}$ we define \begin{eqnarray}
c_{i1}(t)= & \frac{1}{\sqrt{2}}(c_{i+}(t)+c_{i-}(t))\,,\quad c_{i2}(t) & =\frac{1}{\sqrt{2}}(c_{i+}(t)-c_{i-}(t))\label{eq:Keldysh_Rotation1}\\
\bar{c}_{i1}(t)= & \frac{1}{\sqrt{2}}(\bar{c}_{i+}(t)-\bar{c}_{i-}(t))\,,\quad\bar{c}_{i2}(t) & =\frac{1}{\sqrt{2}}(\bar{c}_{i+}(t)+\bar{c}_{i-}(t))\label{eq:Keldysh_Rotation2}\end{eqnarray}
 After this rotation, the action acquires the form \begin{equation}
S=\sum_{a}\int(\overline{c}_{0a}i\partial_{t}c_{0a}+\overline{c}_{1a}i\partial_{t}c_{1a}-H[c])dt-\sum_{a,b}\int C_{ab}(t-t^{\prime})\overline{c}_{1a}(t)c_{1b}(t^{\prime})dtdt^{\prime}\label{TwoSpinAction2}\end{equation}
 where $a,b\in(1,2)$. The Fourier-transform of the matrix $C_{ab}(t-t^{\prime})$
is a standard triangular matrix in Keldysh space: \begin{equation}
\hat{C}(\omega)=\left(\begin{array}{ll}
C^{R}(\omega) & C^{K}(\omega)\\
0 & C^{A}(\omega)\end{array}\right)\ .\label{Cab}\end{equation}
 The functions $C^{R}(\omega)$, $C^{A}(\omega)$ and $C^{K}(\omega)$
are respectively the retarded, advanced and Keldysh component of the
matrix $\hat{C}(\omega)$. If we denote by $\Gamma_{1}(\omega)$ the
spectral density of the external noise $\widehat{h}(t)$ acting on
spin $1$, they are given by \begin{equation}
C^{R}(\omega)=-i\Gamma_{1}(\omega)\,\quad C^{A}(\omega)=i\Gamma_{1}(\omega)\,\quad C^{K}(\omega)=-2i\Gamma_{1}(\omega)\tanh\frac{\omega}{2T}\label{Ca2b}\end{equation}
 Our goal is to determine the spectral density of noise acting on
spin $0$. The bare retarded fermionic Green function at site $1$
(in the absence of the coupling to site $0$) is \begin{equation}
G_{1,(0)}^{R}(\omega)=(\omega-2\xi_{1}+i\Gamma_{1}(\omega))^{-1}\ .\label{GR1}\end{equation}
 The action (\ref{TwoSpinAction2}) is quadratic; diagonalizing it
we get the imaginary part of the retarded fermionic Green function
at site $0$: \begin{equation}
\Im G_{0}^{R}(\omega)=-\frac{(2g/K)^{2}\Gamma_{1}(\omega)}{(\omega-2\epsilon_{1})^{2}(\omega-2\epsilon_{0})^{2}+(\Gamma_{1}(\omega))^{2}(\omega-2\xi_{0})^{2}}\ .\label{eq:ImG^R_0}\end{equation}
 Here $\epsilon_{1,2}=\frac{1}{2}(\xi_{1}+\xi_{2})\pm\sqrt{\frac{1}{4}(\xi_{1}-\xi_{2})^{2}+(2g/K)^{2}}$
are the eigenvalues of the two-spin Hamiltonian (\ref{eq:H_TwoSpin})
in absence of the fluctuating field. In the following we shall neglect
the effects of level repulsion which translates into the difference
between actual energies $\epsilon_{1},\epsilon_{2}$ and their unperturbed
values $\xi_{1},\xi_{2}$. This is the same approximation that we
employed in Sect. \ref{sec:Formation-of-a-long-range} to obtain analytical
results. As we have shown above, this approximation remains accurate
even for very modest values of $K$. Using this approximation we can
rewrite (\ref{eq:ImG^R_0}) as \[
\Im G_{0}^{R}(\omega)=\Im\left\{ \frac{1}{\omega-2\xi_{0}+\frac{(2g/K)^{2}\Gamma_{1}(\omega)}{(\omega-2\xi_{1})^{2}+(\Gamma_{1}(\omega))^{2}}}\right\} \]
 which implies that the imaginary part of the self-energy is \begin{equation}
\Im\left[G_{0}^{R}(\omega)\right]^{-1}=\frac{(2g/K)^{2}\Gamma_{1}(\omega)}{(\omega-2\xi_{1})^{2}+(\Gamma_{1}(\omega))^{2}}\label{eq:ImG^R_0^-1}\end{equation}
 as one expects from perturbation theory arguments.

At zero temperature these results for the fermion Green function are
translated directly into spin correlators. At non-zero temperature
this conversion is less trivial because the spin correlators acquire
an additional decay compared to fermions. Physically, this is due
to the fact that the spin Hamiltonian is non-linear, so the thermal
excitation of one spin might lead to the relaxation of another spin.
For this process to happen the thermal excitation should be mobile,
meaning that it should have an energy larger than $\omega_{d}.$ This
makes such processes rare at low temperatures. Formally, the retarded
transverse spin Green function at site $0$, denoted as $D_{0}^{R}(\omega)$,
differs from the retarded fermion Green function, $G_{0}^{R}(\omega)$,
because the former contains an additional factor $\widetilde{D}_{1}=<\sigma_{1}^{z}\sigma_{1}^{z}>$
due the non-linearity of the Jordan-Wigner transformation. More precisely,
in Keldysh technique these two Green functions are related by \begin{equation}
D_{0}^{R}(t)=\frac{i}{2}\left(G_{0}^{R}(t)\widetilde{D}_{1}^{K}(t)+G_{0}^{K}(t)\widetilde{D}_{1}^{R}(t)\right)\label{DG0}\end{equation}
 The dynamics of the $z$-spin components is purely dissipative in
the leading order in $J$ which we consider here. In this approximation
the retarded Green function of the $z$-spin components is zero; this
allow us to neglect the second term in the general expression (\ref{DG0})
for $D_{0}^{R}(t)$. The first term has two parts: $\frac{i}{2}\widetilde{D}_{1}^{K}=m^{2}+\frac{i}{2}\widehat{D}_{1}^{K}$
where $m=<\sigma_{1}^{z}>=\tanh(\xi_{1}/T)$ and $\frac{i}{2}\widehat{D}_{1}^{K}$
is the irreducible part of the symmetrized spin-spin correlator at
site $1$:\[
i\widehat{D}_{1}^{K}=\left\langle \sigma_{1}^{z}(t)\sigma_{1}^{z}(0)+\sigma_{1}^{z}(0)\sigma_{1}^{z}(t)\right\rangle -2\left\langle \sigma_{1}^{z}(0)\right\rangle \left\langle \sigma_{1}^{z}(t)\right\rangle \]
 Using Eqs.(\ref{JordanWigner}, \ref{GR1}) we get: \begin{equation}
\frac{i}{2}\widehat{D}_{1}^{K}(\omega)=2(1-m^{2})\frac{2\Gamma_{1}(2\xi_{1})}{\omega^{2}+\left[2\Gamma_{1}(2\xi_{1})\right]^{2}}\label{DK1}\end{equation}

In the following we shall be mostly interested in the relaxation at
low energies $\omega\ll\omega_{d}$. Combining Eqs.(\ref{DG0},\ref{DK1})
and (\ref{GR1}), we get the equation for the imaginary part of the
retarded transverse spin-spin correlator at site $0$ at non-zero
temperature: \begin{equation}
-\Im(D_{0}^{+-,R}(\omega))=m_{1}^{2}\frac{\Gamma_{0}^{(0)}(\omega)}{(\omega-2\xi_{0})^{2}}+(1-m_{1}^{2})\frac{\Gamma_{0}^{(0)}(\omega+2i\Gamma_{1}(\xi_{1}))+2\Gamma_{1}(\xi_{1})}{(\omega-2\xi)^{2}+(\Gamma_{0}^{(0)}(\omega+2i\Gamma_{1}(\xi_{1}))+2\Gamma_{1}(\xi_{1}))^{2}}\label{ImD^R0}\end{equation}
 Here $\Gamma_{0}^{(0)}(\omega)$ is the relaxation rate at $T=0$;
it coincides with $\Im\left[G_{0}^{R}(\omega)\right]^{-1}$ given
by Eq.(\ref{eq:ImG^R_0^-1}). Below we neglect the level widths compared
to level energies, which allows to find the imaginary part of the
inverse transverse spin correlation function at site $0$, i.e. the
relaxation rate of the spin at site $0$: \begin{equation}
\Gamma_{0}(\omega)=\Im\left[D_{0}^{R}(\omega)\right]^{-1}=\left(\frac{2g}{K}\right)^{2}\left\{ \frac{m^{2}\Gamma_{1}(\omega)}{\left(\omega-2\xi_{1}\right)^{2}+\Gamma_{1}^{2}(\omega)}+\frac{(1-m^{2})2\Gamma_{1}(2\xi_{1})}{\left(\omega-2\xi_{1}\right)^{2}+4\Gamma_{1}^{2}(2\xi_{1})}\right\} \ .\label{eq:Gamma_0(omega)}\end{equation}
 The second term in this equation is due to the non-linearity discussed
above; in this term we have neglected the contribution of $\Gamma_{1}(\omega)$
compared to $\Gamma_{1}(2\xi_{1})$. The reason for this is that this
term is exponentially small at low $T$ due to the factor $1-m^{2}\ll1.$
Thus, this term makes a significant contribution only if the energy
of spin $1$ is large: $2\xi_{1}>\omega_{d}$ , so that $\Gamma_{1}(2\xi_{1})$
is much larger than $\Gamma_{1}(\omega)$.

The result (\ref{eq:Gamma_0(omega)}) describes the relaxation in
the system of two spins. We now consider a full cavity problem in
which spin $i$ has $K$ neighbors, labeled $k=1,\dots,K$, and each
of these neighbors feels a fluctuating external field with spectral
density $\Gamma_{k}(\omega)$. We compute the relaxation rate of the
spin at site $0$. Adding the contributions from all neighbors we
get\begin{equation}
\Gamma_{i}(\omega)=\left(\frac{2g}{K}\right)^{2}\sum_{k(i)}\left\{ \frac{m^{2}\Gamma_{1}(\omega)}{\left(\omega-2\xi_{1}\right)^{2}+\Gamma_{1}^{2}(\omega)}+\frac{(1-m^{2})2\Gamma_{k}(2\xi_{1})}{\left(\omega-2\xi_{1}\right)^{2}+4\Gamma_{k}^{2}(2\xi_{1})}\right\} \label{eq:Gamma_with_source_term}\end{equation}
 In the zero temperature limit $m_{k}\to1$ the second term in (\ref{eq:Gamma_with_source_term})
vanishes and this formula reduces to (\ref{eq:Gamma_Nonlinear_mapping}).

At non-zero temperatures, one needs to take into account the second
term in (\ref{eq:Gamma_with_source_term}), which becomes important
for low frequencies $\omega<\omega_{d}$ for which $\Gamma_{k}(\omega)$
would be zero at $T=0$. In this regime the second term works as a
source term to the recursion that would otherwise give zero.

\subsubsection{Solution of the recursive equations with a source term and consequences
for low temperature properties.}

At non-zero temperature the low energy modes which are discrete at
$T=0$ acquire a finite lifetime. We shall estimate their broadening
at very low temperature when the effect of non-zero $T$ on modes
with $\omega>\omega_{d}$ can be neglected. Our starting point is
equation (\ref{eq:Gamma_with_source_term}) which contains two terms
of very different physical meaning. The first term describes the decay
due to the indirect coupling to the external fields far away, it is
the only term present at $T=0$. The second term describes the relaxation
caused by mobile thermal excitations of a neighboring spin with energy
above the threshold $\omega_{d}$. At low temperatures such excitations
occur exponentially rarely and thus are essentially limited to a narrow
range of energies slightly above $\omega_{d}$. Because the density
of these excitations is very low, we can ignore the situations in
which a given site feels more than one thermal excitation nearby.
It implies that the relaxation rate induced by these excitations for
low-energy modes is much smaller than the rate of the mobile excitations.
This allows us to replace the effect of the mobile excitations on
low energy modes by an effective fluctuating field. As we shall see
below the spectrum of this field is featureless, so in all respects
the mobile excitations are similar to the external field.

The approximation discussed above allows us to replace the relaxation
rate $\Gamma_{k}(2\xi_{k})$ in the second term in (\ref{eq:Gamma_with_source_term})
by its typical value and to ignore the effect of a non-zero temperature
on this term. Eq. (\ref{eq:Gamma_with_source_term}) becomes very
similar to Eq.(\ref{Bext}) for the susceptibility: \begin{eqnarray}
\Gamma_{i}(\omega) & = & \sum_{k(i)}\frac{\left(2g/K\right)^{2}\Gamma_{k}(\omega)}{\left(\omega-2\xi_{k}\right)^{2}+\Gamma_{k}^{2}}+\eta(T,\omega)\label{eq:Gamma_with_source_term_2}\\
\eta(T,\omega) & \equiv K & \left(\frac{2g}{K}\right)^{2}\int_{\omega_{d}}^{1}\frac{dx}{\cosh^{2}(x/2T)}\,\frac{\Gamma_{typ}(x)}{(x-\omega)^{2}+4\Gamma_{typ}^{2}(x)}\nonumber \end{eqnarray}
 where $\Gamma_{typ}(x)$ is given by the zero-temperature result,
see Eq.(\ref{eq:Gamma^typical(omega)}). The dominant contribution
to the external dissipation $\eta(T,\omega)$ in Eq.(\ref{eq:Gamma_with_source_term_2})
is determined by the competition between $\exp(-x/T)$ and $\Gamma_{typ}(x)\approx e^{-1/eg}\exp[-\omega_{1}/(x-\omega_{d})]$.
Evaluating the integral over $x$ in the saddle point approximation
(valid under the conditions $\omega<\omega_{d}$ and $T\ll\omega_{1}$
) we find that $\eta(T,\omega)$ is weakly $\omega$-dependent and
given by \begin{equation}
\eta(T,\omega)=C\frac{4g^{2}}{K}\frac{4\sqrt{\pi}\omega_{1}e^{-1/eg}}{(\omega_{d}+\sqrt{\omega_{1}T}-\omega)^{2}}\left(\frac{T}{\omega_{1}}\right)^{3/4}\exp\left[-\frac{\omega_{d}}{T}-2\sqrt{\frac{\omega_{1}}{T}}\right]\label{eq:eta(T)}\end{equation}
 The estimate (\ref{eq:eta(T)}) is valid with exponential accuracy
provided that $T\ll\omega_{1}$. The weakness of the frequency dependence
in $\eta(T,\omega)\approx\eta(T)$ proves our statement above according
to which the noise produced by the thermal excitations is featureless
and exponentially small.

Using the formal analogy between Eq.(\ref{eq:Gamma_with_source_term_2})
and the Eq. (\ref{b_i}) for the susceptibility, we can now use the
results obtained in Sec.\ref{sub:Susceptibility-in-the}. The key
conclusion of this section that we need here is that the typical value
of the susceptibility does not contain any divergence at the $T=0$
transition point. This implies that the typical sub-threshold level
width is $\Gamma^{typ}(\omega<\omega_{d})\approx\eta(T)$. This confirms
our conjecture above that the relaxation rate of the low energy modes
is very low, which allowed us to replace the mobile excitations by
an effective fluctuating field. To estimate the level width for the
whole range of frequencies, we note that both $\eta(T)$ and $\Gamma_{0}(\omega)$
(see Eq.(\ref{eq:Gamma^typical(omega)})) are exponentially fast functions,
therefore \begin{equation}
\Gamma^{typ}(\omega,T)\approx\max\left(\eta(T),\Gamma_{0}(\omega)\right)\ .\label{eq:Gamma^typ(omega,T)}\end{equation}
 The estimates (\ref{eq:Gamma^typical(omega)}) and (\ref{eq:eta(T)})
are not valid right at the critical point $g=g_{c}$, where $\omega_{d}$
and $\omega_{1}$ vanish and the level width is given by Eq.(\ref{eq:Gamma_at_Kc}).
Repeating the calculations similar to those used to derive Eq.(\ref{eq:Gamma_at_Kc})
we find at $g\ll1$, instead of (\ref{eq:eta(T)}): \begin{equation}
\eta(T)\propto\exp\left[-\frac{C}{e^{2}g}\left(\frac{2e}{T}\right)^{\frac{eg}{1+eg}}\right]\label{eq:eta(T)_critical}\end{equation}

The equations (\ref{eq:eta(T)},\ref{eq:eta(T)_critical}) give the
typical relaxation rate of low energy modes. At the same time they
give a typical noise level and thus a typical rate of transport processes
in this model. For the superconductor-insulator transition the result
(\ref{eq:eta(T)}) implies that, away from the critical point, the
resistivity follows the Arrhenius law at very low temperatures and
a $\exp(1/\sqrt{T})$ law in the intermediate temperature regime.
This behavior is exactly opposite to the one expected and observed
in conventional Mott insulators where Arrhenius is followed by Mott
behavior as temperature is decreased. Exactly at the critical point,
the resistivity grows as a stretched exponential $\exp(T^{-\alpha})$
with $\alpha\approx0.25-0.3$ for a physically relevant $g\approx0.12$.

\section{The effect of frustration.\label{sec:The-effect-of-frustration}}

Disordered superconducting films close to the SI transition can be
driven to insulators by the application of a magnetic field. The properties
of these materials in presence of a magnetic field have been extensively
studied experimentally and are rather unusual. So it is important
to discuss the theoretical expectations. We will not undertake a full
quantitative study here, but we will only study the leading effects
of a magnetic field on the phase diagram of our model formulated on
the Bethe lattice. In the framework of this model the effect of a
magnetic field is described by the effective model where nearest-neighbor
couplings $M_{ij}$ defined in Eq.(\ref{H_A}) acquire random phases:
$M_{ij}=\frac{g}{Z-1}e^{i\alpha}$, with $\langle\alpha^{2}\rangle=f$.
The effects of these random phases on the two major lines of our phase
diagram (see Figs. \ref{fig:PhaseDiagram} and 8),
the critical temperature line $T_{c}(K)$ and the threshold energy
line in the insulating phase $\omega_{d}(K)$, are crucially different.
Whereas all the equations for level widths $\Gamma_{i}$ contain squares
of absolute values of matrix elements $|M_{ij}|^{2}$ only, and thus
do not depend on $f$, the equations for the order parameter are affected
by the random phases.

Consider first the case of small phase fluctuations $f\ll1$. In the
limit of very large $K>K^{RSB}(g)$ the simple mean-field approximation
should be valid, and the transition temperature is determined by the
equation for the first moment of the $P(B)$ distribution. Random
phases enter this equation via a straightforward modification of the
coupling strength, $g\to g(1-f/2)$, which leads to the suppression
of $T_{c}$: \begin{equation}
\ln\frac{T_{c}(f)}{T_{c}}=-\frac{f}{2g}\label{f1}\end{equation}
 Because $g$ is small, the suppression of the transition temperature
(and therefore the decrease of the typical order parameter in the
ordered phase) can be rather strong even at $f\ll1$.

The case of strong frustration is more complicated. Here we consider
the limit of very large fields that generate completely random phases
$\alpha_{i}$ with uniform distribution over $(0,2\pi)$. Instead
of Eq.(\ref{eq:mapping_Kfinite}) we get \begin{equation}
B_{i}=\frac{g}{K}\sum_{k=1}^{K}\frac{B_{k}e^{i\alpha_{k}}}{\sqrt{B_{k}^{2}+\xi_{k}^{2}}}\tanh\beta\sqrt{B_{k}^{2}+\xi_{k}^{2}}\ .\label{f2}\end{equation}
 and we look for a solution for the $P(B)$ distribution function
which depends only on the absolute value $|B|$.

In order to determine $T_{c}$ we use the linearized version of (\ref{f2})
which can be rewritten in terms of the Fourier transformed distribution
$Q(s,s^{*})=\int dBdB^{*}P(B,B^{*})e^{\frac{i}{2}(sB^{*}+s^{*}B)}$.
We shall assume that this Fourier transform depends only on the absolute
value $|s|$: $Q(s,s^{*})\equiv\hat{Q}(|s|)$. It then satisfies:
\begin{equation}
\hat{Q}(|s|)=\left[\int_{0}^{1}d\xi\hat{Q}\left(|s|\frac{g}{K}\frac{\tanh\beta\xi}{\xi}\right)\right]^{K}.\label{eq:Laplace_transform_field}\end{equation}
 Note that these equations are formally identical to those obtained
in the case without magnetic field in (\ref{eq:Laplace_recursion}).
However, the analytic properties expected in the present case are
different: the random phases induce a symmetric $P(B)$ distribution,
instead of the distribution supported on positive $B$'s when there
is no magnetic field. We will look for a power-law solution of Eq.
(\ref{eq:Laplace_transform_field}) at small $s$ in the form $\hat{Q}(s)=1-A|s|^{x}$.

Because the average order parameter is zero in the limit of large
magnetic fields, the simple mean-field solution is obtained by assuming
that $P(B)$ is determined by its second moment and correspondingly
$x=2$. The self-consistent equation (\ref{eq:Laplace_transform_field})
then gives $T_{c}^{MF}=1.705g^{2}/K$. This is the equivalent of (\ref{gc_naive})
in the zero magnetic field case.

The equation for the Laplace transform (\ref{eq:Laplace_transform_field})
is formally identical to the one obtained in zero field (\ref{eq:Laplace_recursion}),
so the value of the exponent $x$ that determines the behavior of
the Fourier transform at small $s$ is determined by the same equations
(\ref{eq:Tc_nearKc1},\ref{eq:Tc_nearKc2}) derived in Sect. \ref{sub:Phase-diagram}.
The important difference between the cases of zero and large magnetic
field is due to the fact that in the latter case the simple mean field
solution corresponds to $x=2$$.$ As a result, in contrast to zero
field case, the simple mean field solution is not valid even in the
$K\to\infty$ limit. In other terms, RSB \textit{always} occurs for
the fully-random problem defined by Eq. (\ref{f2}).

We begin by solving these equations in the large $K$ limit, where
the exponent $x$ approaches $1+\epsilon$, with $\epsilon\ll1$ as
long as $g\ll1$. Assuming that $\epsilon\ln\frac{1}{T}\gg1$, we
can extend the integrals over $\xi/T$ in Eqs.(\ref{eq:Tc_nearKc1},\ref{eq:Tc_nearKc2})
up to $\infty$. We estimate the resulting integrals: \begin{eqnarray*}
\int_{0}^{\infty}dx\left(\frac{\tanh x}{x}\right)^{1+\epsilon} & \approx & \frac{1}{\epsilon}\\
\int_{0}^{\infty}dx\left(\frac{\tanh x}{x}\right)^{1+\epsilon}\ln\frac{\tanh x}{x} & \approx- & \frac{1}{\epsilon^{2}}\end{eqnarray*}
 and obtain \begin{eqnarray}
\epsilon & = & eg\label{eq:epsilon_frustrated}\\
T_{c} & = & \frac{g}{K}e^{-1/eg}\label{eq:T_frustrated}\end{eqnarray}
 Similarly to the zero magnetic field case, in the RSB phase the naive
mean field prediction $T_{c}^{MF}=1.705g^{2}/K$ is exponentially
larger than the correct result (\ref{eq:T_frustrated}) for small
$g$.

We now prove that for any $K$ the transition occurs in a RSB phase.
To find the temperature, $T_{RSB},$ corresponding to replica symmetry
breaking we consider the Eq. (\ref{eq:Tc_nearKc2}) and assume that
$x=2$, using a procedure similar to the determination of the RSB
point, Eq.(\ref{eq:K_RSB}), discussed in Sec.\ref{sub:Phase-diagram}
for the unfrustrated case. However, in contrast to the zero-field
case, the corresponding temperature $T_{RSB}\approx g/K$, is always
\textit{above} the simple mean field value $T_{c}^{MF}$ at $g\ll1$.
Thus, it is necessary to solve both equations (\ref{eq:Tc_nearKc1},\ref{eq:Tc_nearKc2})
together to determine the value of the exponent $x<2$ and transition
temperature.

\begin{figure}[ht]
 \includegraphics[width=4in]{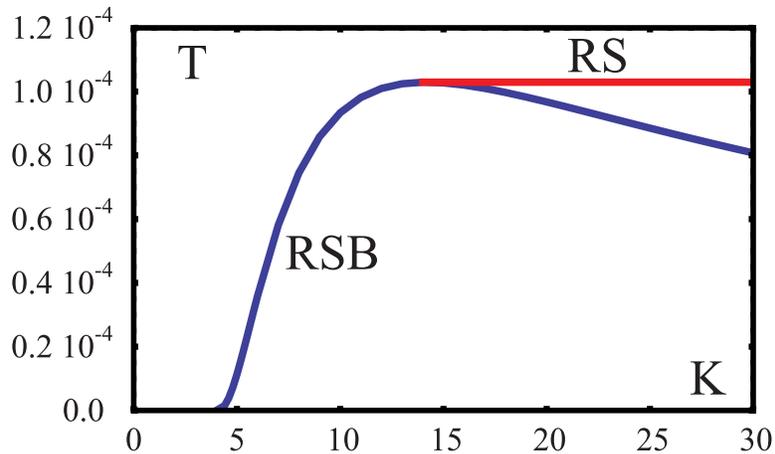} \caption{Blue line: dependence of the critical temperature $T_{c}(K)$ on $K$
of the fully frustrated system for $g=0.1$. Red line: result of the
simple mean field approximation for the problem without frustration.
The maximum of the blue line corresponds to the case where the order
parameter distribution decays with an exponent $x=1$ and $K=K^{RSB}$,
where $K^{RSB}$ is determined for the unfrustrated model by Eq. (\ref{eq:K_RSB}).
At $K<K^{RSB}$ the critical temperatures of the regular and the frustrated
model coincide.}

\label{Fig:PhaseDiagram_LargeField} 
\end{figure}

The applicability of the solution (\ref{eq:epsilon_frustrated},\ref{eq:T_frustrated})
is limited to the regime of very large $K$, when $T$ is so small
that the corrections of order $T^{eg}$ due to the finite
upper limit in the integral $\int_{0}^{1/T}\left(\frac{\tanh x}{x}\right)^{1+eg}$
are negligible. When $K$ decreases these corrections become significant
and the exponent $x$ starts to decrease; it eventually approaches
unity at the value $K=K^{RSB}$ determined in Eq.(\ref{eq:K_RSB}).
At the same time, $T_{c}(K)$ deviates from the simple law (\ref{eq:epsilon_frustrated})
and attains at $K=K^{RSB}$ its maximum value, equal to the mean-field
transition temperature $T_{c0}$ of the unfrustrated model. At still
smaller $K$, the solution for $T_{c}$ is identical to the one discussed
in Sec.\ref{sub:Phase-diagram} for the unfrustrated model. This behavior
is summarized in Fig.\ref{Fig:PhaseDiagram_LargeField}.%

\begin{figure}[ht]
 \includegraphics[width=7cm]{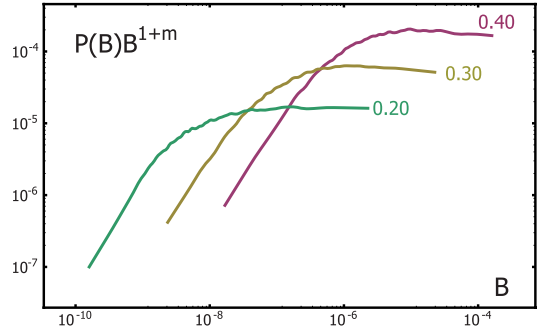} \caption{Distribution function of the local order parameter for the strongly
frustrated model with $K=4$ at three different values of the coupling
constant. The number near each curve indicates the corresponding value
of $({g}/{g_{c}}-1)$. The plateau in $P(B)B^{m+1}$ demonstrates
the existence of the scaling regime (\ref{eq:P(B)_scaling_form}). }

\label{Fig:P(H)_K=00003D4_RandomPhases} 
\end{figure}

\begin{figure}[ht]
 \includegraphics[width=7cm]{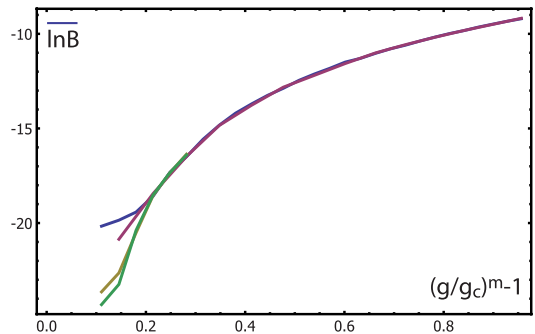} \caption{Typical amplitude of the order parameter as function of the proximity
to the quantum critical point.}

\label{Fig:LogB_K=00003D4_RandomPhases} 
\end{figure}

Although the transition line $T_{c}(K)$ stays the same (in the RSB
phase) for the unfrustrated model and for the strongly frustrated
one, the amplitudes of the order parameter in the ordered phase differ
considerably. This is illustrated by the numerical solution of the
random-phase mapping equations (\ref{f2}) shown in Figs. \ref{Fig:P(H)_K=00003D4_RandomPhases}
and \ref{Fig:LogB_K=00003D4_RandomPhases} for zero temperature and
near to the quantum critical point $g=g_{c}$.

The results shown in Fig. \ref{Fig:LogB_K=00003D4_RandomPhases} are
very much like those seen in Fig. \ref{fig:LogHtyp_K=00003D4} for
the unfrustrated model, but the typical amplitude $\overline{B^{typ}}$
is suppressed by a factor $\sim100$ in the frustrated case at the
same $g/g_{c}$ value.

To summarize this section, we have demonstrated that frustration suppresses
strongly the transition temperature $T_{c}$ when one is in the replica
symmetric phase at sufficiently large $K$ but it has no effect on
$T_{c}$ in the RSB phase near the quantum phase transition. This
can be interpreted as a consequence of the quasi-one-dimensional nature
of Bethe-lattice clusters which contribute to the formation of the
coherent state in the RSB phase. While $T_{c}$ is unchanged in the
RSB region, the amplitude of the order parameter (at $T\ll T_{c}$)
is strongly suppressed due to frustration, as demonstrated in Fig.~\ref{Fig:LogB_K=00003D4_RandomPhases}.
More work is needed to decide if these results, obtained on the Bethe-lattice
problem, can be applied to the finite-dimensional problem where closed
loops are present (see discussion in section \ref{sub:The-effect-of-a-magnetic}).
We expect, however, that the results will remain qualitatively similar
due to the dominance of a small number of paths that makes the presence
of small loops largely irrelevant.

\section{Consequences for experiments.\label{sec:Consequences-for-experiments.}}

\subsection{Distribution of coherence-peak heights in STM and Andreev point contact
tunneling experiments. \label{sub:Distribution-of-coherence}}

One of the main results of this work is anomalous broadening of the
distribution of the local values of the order parameters in the vicinity
of the SI transition. This conclusion can be tested by STM measurements.

The same STM experiments can also confirm that the transition happens
due to Cooper pair localization and not by unbinding of Cooper pairs.
In a usual setup of these experiments one measures the current-voltage
characteristic of a highly resistive tunneling contact between the
material and a needle of the STM machine. The measured differential
conductance, $dI/dV$, is proportional to the single-electron density
of states, $\rho(E)$, at energy $E=eV$. Thus, these measurement
directly probe the superconducting gap. The experiments performed
on superconductor-insulator transition in TiN \cite{Sacepe2007,Sacepe2009}
and InO \cite{Sacepe2007,Sacepe2010} demonstrate that in these materials
the superconductor-insulator transition happens without pair destruction
(the single-electron gap remains large). The theoretical justification
of this was given in the work\cite{Feigelman2010} and is summarized
above in section \ref{sub:Experimenatal-results}. Thus, one expects
that this transition is driven by the competition between disorder
and Cooper-pair tunneling and is described by the models studied in
this paper. As we discussed in section \ref{sub:Experimenatal-results}
the present paper ignores subtle effects of the correlations between
matrix elements, in this approximation the disorder changes only the
number of neighbors, $K$. In order to compare with experimental data
we need to choose the interaction constant so that to reproduce the
correct value of $ T_{c}/E_F \approx10^{-3}$ away from the transition
in these materials. This gives the interaction $g=0.129$ used in
many numerical plots in the previous sections. 

In superconductors the density of states above the gap displays the
coherence peak which is due to the superconducting order parameter.
The height's statistics of these peaks can be used to measure the variations of the
local order parameter appearing in the vicinity of the superconductor-insulator
transition. In particular, the broadening of the $P(B)$ distribution
translates into the broad distribution of peak heights. This theoretical
prediction was indeed confirmed in very recent experiments \cite{Sacepe2010}
where superconducting samples with different values of $T_{c}$ were
studied and compared. The data shows that while the superconducting
gap remains largely intact in lower $T_{c}$ samples and experience
modest fluctuations from point to point, the distribution of peak
heights changes dramatically as the insulator is approached. This
observation is in full agreement with the theory developed in this
work, and confirms our expectation that the solution of the model
on Bethe lattice provides a  good approximation for the properties
of realistic physical systems.

Another possible set of data is given by Andreev point-contact tunneling
measurements. Similar to conventional tunneling, these experiments
give current-voltage characteristics of the point contact. However,
in contrast to conventional contacts, the contacts used in these experiments
are characterized by a small resistance, $R\sim1k\Omega$. This allows
a coherent tunneling of the Cooper pair between the normal needle
and a superconducting material.

In a conventional BCS superconductor the differential conductivity
observed in Andreev point-contact experiments shows a suppression
below the single-particle gap and a peak at the gap edge, similar
to the conventional tunneling. The reason for this is that in conventional
superconductors the gap, $\delta E_{pair}$, for collective pair excitations
(pseudospins) is exactly twice as large as the gap for single particle
excitations: $\delta E_{pair}=2\Delta_{BCS}$. Tunneling of Cooper
pair brings energy $2eV$, so the gap edge observed in differential
conductivity for the single particle tunneling coincide with the one
observed in Cooper pair tunneling.

In contrast to conventional superconductors, strongly disordered superconductors
in the vicinity of the SI transition are expected to show much smaller
gap for collective pair excitations, $\delta E_{pair}$, than the
one characterizing single-particle tunneling, $\Delta_{sp}$. In this
case, the observed differential conductivity below single-particle
gap is due to the coherent Cooper pair tunneling which might occur
at potential differences $eV<\Delta_{sp}$. In the lowest non-zero
order in tunneling one gets that at such bias the differential conductivity
is proportional to the local susceptibility, $\mbox{Im}\chi_{r}(\omega=eV)$,
of the equivalent spin model evaluated at the position of the needle,
$r$. This expectation is correct provided that the collective modes
contributing to the local susceptibility are delocalized at this frequency.
The tunneling into localized modes cannot lead to experimentally observable
conductivity. Because the energy separating the localized and delocalized
modes is the same at different positions in the sample, we expect
that the observed differential conductivity is characterized by a
spatially uniform threshold or a peak. This expectation is in agreement
with the preliminary results \cite{Dubouchet2011}.

\subsection{Low-temperature resistivity in the insulating state\label{sub:Low-temperature-resistivity-in}}

Because STM measurements require that the resistance of the tunneling
contact is much larger than the resistance of the sample, they become
difficult, if not impossible, in the insulating phase. So, in this
state one should compare the theoretical predictions with the transport
data which are always more difficult to interpret. The most striking
feature of the resistivity on the insulating side of the transition
in InO and TiN is the activated behavior of the resistance $R\sim\exp(T_{A}/T)$,
with an activation energy that decreases in the vicinity of the transition.
\cite{SacepeShahar2009,Sacepe2008} This decrease of the activation
energy, $T_{A}$, is in contrast with a large and non-critical single
particle gap observed in STM \cite{Sacepe2007,Sacepe2009,Sacepe2010}
on the superconducting side of the transition. It indicates that the
transport in this regime is due to incoherent Cooper pair hopping.
More direct evidence of the pair transport is provided by the data
on magnetoresistance oscillations in perforated films which displays
periodic oscillations with {}``superconducting'' flux period $\Phi_{0}=hc/2e$
in the \emph{insulating} state.\cite{Valles2009}

As we have shown in Section \ref{sub:Propagation-of-time-dependent},
the low lying excitations of the spin model have zero width. This
translates into the localization of all excitations with energies
$E<\omega_{d}$. The transport is made possible by modes with energies
$E>\omega_{d}$ which become thermally excited at non-zero temperatures.
Thus, we expect that the resistivity in this state has an approximately
activated temperature dependence at lowest temperatures with an activation
energy $T_{A}=\omega_{d}$ which decreases when one approaches the
transition. The subdominant term in the resistance is proportional
to $\exp\sqrt{\omega_{1}/T}$ which might become important at intermediate
temperatures because $\omega_{1}\gg\omega_{d}$. This result is in
agreement with the data. It should be contrasted with the naive expectation
according to which the transport of Cooper pairs could be characterized
by Mott (or Efros-Shklovskii) behavior, giving a behavior $R\sim\exp(T_{M}/T)^{a}$
($a=0.25-0.5),$ in analogy with single electron transport in conventional
insulators. In these insulators the activation behavior is realized
at intermediate temperature range and Mott behavior at lowest temperatures.

The sharp boundary, $\omega_{d}$, separating the localized and delocalized
modes could be more directly probed by microwave experiments. We expect
that close to this boundary the lifetime of the excitations becomes
very long. Thus, even a weak microwave radiation could excite a large
number of these excitations. This would dramatically increase the
conductivity of the sample without affecting significantly its temperature.
Furthermore, applying the microwave power in pulses with varying time
intervals between the pulses, one could obtain information on the
lifetime of these excitations near the many-body mobility edge.

\subsection{The effect of frustration induced by a magnetic field or by unconventional
order parameter symmetry. \label{sub:The-effect-of-a-magnetic}}

As discussed in Section~\ref{sec:The-effect-of-frustration}, the
highly inhomogeneous spatial structure appearing in the vicinity of
the SI transition strongly suppresses the effects of frustration and
thus of a magnetic field. In the approximation used in this work the
magnetic field has no effect on the transition temperature in this
phase. The account of higher order effects in $1/K$ would lead to
a small suppression of the critical temperature in this regime. This
conclusion is in qualitative agreement with the data: it has been
found (see e.g. \cite{Sacepe2007,Sacepe2008,Sacepe2009,Sacepe2010}
and references therein) that the properties of the materials depend
very sensitively on the disorder level in the vicinity of the SI transition
in contrast to their smooth dependence on a magnetic field. In spite
of its weak effect on the transition temperature, we have observed
that the magnetic field should affect strongly the order parameter.
This prediction can be tested by STM and Andreev point-contact measurements
which provide direct information on the local superconducting order
parameter and its distribution.

The very weak effect of a magnetic field on the transition temperature
implies that superconductivity might survive in strongly disordered
superconductors with unconventional pairing symmetries such as d-wave
symmetry, provided that the Cooper pairs in these superconductors
are formed on very short scales which are weakly affected by disorder.
This might be the case in high $T_{c}$ superconductors in the pseudogapped
regime in which there are reasons to expect\cite{Geshkenbein1997,Galitski2009}
that electrons are paired in the corners of the Brillouin zone. In
this case the Cooper pairs have very small size and are not affected
by disorder. The global superconductivity is due to the coupling between
these localized Cooper pairs. These couplings have random signs due
to the d-wave symmetry of the order parameter which makes this situation
similar to the strongly disordered superconductor in a magnetic field.
As discussed above the randomness of the couplings has weak effect
on the transition temperature. This might explain the apparent similarity
between underdoped high $T_{c}$ superconductors and strongly disordered
InO and TiN films, noted by many researchers previously, see for example~\cite{Steiner2005a,Steiner2005b}

\subsection{Change of level statistics: suggestion for numerical work. \label{sub:Change-of-level}}

The many-body mobility edge line $\omega_{d}(K,g)$ found in Section
\ref{sub:Propagation-of-time-dependent} divides the parameter space
of the Hamiltonian into regions with qualitatively different spectral
statistics. At $\omega<\omega_{d}$ the spectrum is point-like and
the many-body wavefunctions are localized, thus we expect the level
statistics to be Poissonian in this region. On the other hand, above
the mobility edge, at $\omega>\omega_{d}$, a nonzero line width is
generated, indicating the extended nature of wavefunctions. In this
regime we expect a Wigner-Dyson level statistics, with level repulsion.
This change of level statistics can be studied numerically via exact
diagonalization of the Hamiltonian (\ref{H_A}).

A similar study was implemented recently in the work\cite{Huse2007}
for a one-dimensional model of interacting fermions. This paper has
found a qualitative evolution between Poissonian and Wigner-Dyson
statistics, but large finite-size effects make any quantitative conclusions
difficult in the absence of an appropriate analytic theory. We expect
that the Hamiltonian (\ref{H_A}) with moderate branching number $K>1$
could be useful for such study, because our theoretical predictions
for the position of the mobility edge can be used for comparison with
the numerical data.

As a next step, it would be important to extend such a numerical study
to the same type of quantum spin models on a usual real-space lattice.
The comparison of the results for spectral statistics obtained on
conventional Euclidean lattices and on the Bethe lattice could be
very useful to check the applicability of the approach developed in
this paper to physical systems.

\section{Conclusions \label{sec:Conclusion}}

In conclusion, we have outlined a solution of the strongly disordered
spin model on the Bethe lattice that can be mapped onto the disorder-driven
superconductor-insulator transition or onto the disordered ferromagnets.
We found a series of two zero temperature transitions between an ordered
state, a state with a slow relaxation and a state with no relaxation.
Our solution also shows that the low temperature phase in this model
is always very strongly non-uniform with both the order parameter
formation and the spin relaxation controlled by rare paths that contain
a very small fraction of spins. When applied to the superconductor-insulator
transition our results imply the existence of both weak and strong
insulators. In the former the relaxation rate varies faster, as $\exp(\sqrt{\omega_{1}/T})$,
in an intermediate temperature range, but it crosses over to Arrhenius
$\exp(\omega_{d}/T)$ at very low temperatures. Exactly at the quantum
critical point we expect a relaxation rate that varies as $\exp(1/T)^{\alpha}$
with a non-universal $\alpha$ that depends on the interaction constant.
For physically relevant values of the interaction constant we expect
$\alpha\approx0.25-0.3$.

In the strong insulator the relaxation is completely suppressed. Of
course, the physical effects neglected in our model (like electron-phonon
coupling) would lead to some slow relaxation even in the latter phase.

The predictions of our theory can be tested by conventional STM experiments
and Andreev point-contact tunneling experiments on superconducting
films of InO and TiN as discussed in detail in Section \ref{sec:Consequences-for-experiments.}.

In the strongly insulating phase the local quantum levels remain discrete,
this implies that such systems would remain coherent for a very long
time (limited by the interaction with phonons and other environment
neglected in our model). When applied to disordered magnets this result
implies that a stronger disorder might make the spin system less noisy
and more coherent. This might be important for the efforts to suppress
the flux noise in superconducting circuits where it is believed\cite{Faoro2008,Koch2007}
to be due to the disordered spins at the surface of the superconductors.

Our theory is directly applicable to the models on Bethe lattices
which do not have small loops. Qualitatively it looks very likely
that the dominance of a small number of paths, which is a central
result of this work, implies that small loops are irrelevant and therefore
the Bethe-lattice approximation should be very good. This conjecture
might be checked numerically (see Section \ref{sub:Change-of-level}).

We now briefly discuss possible extensions of the present theory and
list few open problems: 
\begin{itemize}
\item Extension of the developed technique to the spin lattices in Euclidean
space. It is believed that the exponential behavior of physical properties
in the localization problem on a Bethe lattice disappears in any finite
dimension\cite{Mirlin1994}. However, the arguments of \cite{Mirlin1994}
are not directly applicable to the non-linear problem of phase transition
that we consider here in which the phase-space dimension increases
exponentially with the number of sites. Qualitatively it seems likely
that the dominance of a small number of paths is an indication that
the main results of the Bethe-lattice approximation will remain correct
in finite dimensions. 
\item The extension to finite-dimensional systems would be greatly simplified
if one could find the appropriate order parameters for the phase transitions
found on Bethe lattice. In particular, these order parameters should
describe not only the appearance of the superconducting order but
also its homogeneity and its time-reversal symmetry breaking which
appears at the second transition. 
\item Direct calculation of transport quantities such as thermal conductivity
or electrical resistivity near the quantum transition studied in this
work. The present paper computes the spin relaxation time which for
the SI problem translates into the time required for the Cooper pair
to move away from a given site. With exponential accuracy it should
be the same as the resistivity or thermal conductivity in the system.
In order to compute the properties with better accuracy one would
need to develop the formalism to compute transport properties directly
in this problem. 
\item Direct computation of the dynamic spin susceptibility $\chi_{r}(\omega)$.
The behavior of this quantity determines the pair conductivity measured
in STM experiments and Andreev contact spectroscopy. 
\item Finite temperature effects in the strong insulator. In the approximation
used in this paper even a high temperature has no effect on the strong
insulator. However, one might worry that our approximation might miss
exponentially small effects such as rare regions characterized by
larger density of states that would make this part of the system less
insulating. 
\end{itemize}

\section{Acknowledgments}

We acknowledge useful discussions with G. Biroli, C. Chapelier, T.
Dubouchet, V. Kravtsov, M. Mueller, F. Zamponi and B. Sacepe. This
work was supported by Triangle de la physique 2007-36, ANR-06-BLAN-0218,
ECS-0608842, ARO 56446-PH-QC, DARPA HR0011-09-1-0009 and by the program
\textquotedbl{}Quantum physics of condensed matter\textquotedbl{}
of the Russian Academy of Sciences, and RFBR grant 10-02-00554.

\section{Appendix A: Quantum cavity method: Suzuki-Trotter formulation and
further approximations }

If one is interested only in the thermodynamic properties of the problem
described by (\ref{H_A},\ref{H_B}) on the Bethe lattice (a random graph of connectivity
$Z=K+1$), one can write \cite{Laumann} an exact (at the replica symmetric
level) cavity recursion in terms of Suzuki-Trotter paths. Unfortunately
this exact recursion can be studied only numerically and is rather
heavy to handle in the case of disordered systems. Our approach uses
a simplified version which amounts to using a special measure on the
Suzuki-Trotter paths. In this appendix we briefly summarize the exact
cavity recursion and we make explicit the extra approximations of
our method. Let us study for instance the transverse-field Ising Hamiltonian
(\ref{H_B}). \begin{equation}
H_{I}=-\sum_{i}\xi_{i}\sigma_{i}^{z}-\frac{g}{Z-1}\sum_{(ij)}\sigma_{i}^{x}\sigma_{j}^{x}\label{H_B1}\end{equation}

The partition function $Z=Tr\ e^{-\beta H_{I}}$can be expressed with
the Suzuki-Trotter representation. Using $N_{t}$ imaginary time steps,
one introduces the time trajectory of each spin, $\sigma_{i}(t)\in{\pm1}$
, where $t=1,\dots,N_{t}$. Then \begin{equation}
Z=\lim\ \sum_{N_{t}\to\infty\left\{ \sigma_{i}(t)\right\} }e^{-\beta H_{ST}}\end{equation}

where:\begin{equation}
H_{ST}=-\frac{1}{N_{t}}\sum_{t}\sum_{(ij)}\frac{g}{Z-1}\sigma_{i}(t)\sigma_{j}(t)-\sum_{t}\sum_{i}\Gamma_{i}\sigma_{i}(t)\sigma_{i}(t+1)\end{equation}

and $\Gamma_{i}=\frac{1}{\beta}\log\cosh\left(\frac{\beta\xi_{i}}{N_{t}}\right)$

The exact RS method introduced in \cite{Laumann} is the following:
take a branch of the Bethe lattice rooted on spin $i$, and define
the probability distribution of the time-trajectory of this spin as
$\psi_{i}\left[\sigma_{i}(t)\right]$. Let us define analogously,
for each of the $K$ spins $ $j which are the first neighbors of
$ $i on the rooted tree, their time trajectory as $\psi_{j}\left[\sigma_{j}(t)\right]$.
Then one can write formally the mapping that generates, from $\left\{ \psi_{j}\left[\sigma_{j}(t)\right]\right\} $
 the new $\psi_{i}\left[\sigma_{i}(t)\right]$. This mapping is
then typically studied numerically by a population dynamics method
where one represents each of the $\psi_{j},\ \psi_{i}$ by a sample
of spin trajectories. In paper~\cite{Krzakala} it is shown how to take the
$N_{t}=\infty$ limit by using a continuous time formulation where
one memorizes only the times at which the spin trajectories jump.
The whole procedure can be seen as a two-step approach: transforming
the problem to one of classical spin trajectories, and then using
the classical cavity method \cite{MezardParisi2001}. In particular, the
cavity mapping can be shown to derive from a variational principle
based on a Bethe free energy.

The quantum cavity approach that we have introduced in Sect. \ref{sub:Cavity-mapping} can
be understood as an approximate way to study the exact RS Suzuki-Trotter
based cavity method. Basically, the hypothesis amounting to using
a local Hamiltonian of spin $k$ in the form $\xi_{k}\sigma_{k}^{z}+B_{k}\sigma_{k}^{x}$
amounts, in the Suzuki-Trotter formalism, to using a local distribution
$\psi_{k}\left[\sigma_{k}(t)\right]$ which takes the special form
\begin{equation}
\psi_{k}\left[\sigma_{k}\left(t\right)\right]=C\exp\left(\frac{\beta B_{k}}{N_{t}}\sum_{t}\sigma_{k}(t)+\beta\Gamma_{k}\sum_{t}\sigma_{k}(t)\sigma_{k}(t+1)\right)\end{equation}
 This form can be used with different methods. In a variational approach
one injects this form into the Bethe free energy. This gives a free
energy which depends on all the parameters $B_{k}$, on which one
should optimize. In practice this method is a bit heavy numerically
because the computation of this Bethe free energy as function of the
$B_{k}$ involves, for each spin $i$, a trace over $\sigma_{i}$ and
all its neighbors $\sigma_{k}$, which is thus a sum over $2^{K+2}$
terms. In the approach that we have described in Sect.~\ref{sub:Cavity-mapping}  we use a slightly
different method: we suppose that all the cavity neighbors of a given
spin $i$ have distribution $\psi_{k}\left[\sigma_{k}(t)\right]$.
 From these distributions we compute $\langle\sigma_{i}^{x}\rangle$
and deduce from it the value $B_{i}$ which reproduces this average.
This defines a mapping from $\left\{ B_{k}\right\} $ to $B_{i}$.
This type of approach can be tested by comparing it to the full numerical
sampling of Suzuki-Trotter trajectories in the case of a uniform transverse
field where $\xi_{i}=\xi$. This study, done in \cite{Zamponi},
confirms that this approach is able to reproduce the phase diagram
accurately (the variational approach is a bit more precise, but both
approaches give the zero-temperature value of $g_{c}$ within a few
per-cent).

Our approach uses one more step of approximation: instead of the quantum
cavity mapping described above, we have mostly used the explicit
mapping
(\ref{eq:mapping_Kfinite})
 which has the advantage that one does not need to compute $\langle\sigma_{i}^{x}\rangle$
in order to find the value of $B_{i}$. The validity of this further
step of approximation has been studied in
Sect.\ref{sub:Leading-correction-in}.

\end{document}